\documentclass[journal]{./IEEEtran}

\usepackage{amsfonts} 

\usepackage{array}
\usepackage{lipsum}

\usepackage{xspace}
\usepackage{algorithm}
\usepackage[noend]{algpseudocode}
\usepackage{amsmath}
\usepackage{subcaption}
\usepackage{relsize}
\usepackage{xcolor}

\newcommand{\BigO}[0]{\mathcal{O}}
\newcommand\bigforall{\mbox{\Large $\mathsurround=0pt\forall$}}

\usepackage{multirow}

\newcommand{\GDS}{GDS\xspace}

\newcommand{\APPR}{HUDD\xspace}

\newcommand{\CHANGED}[1]{\textcolor{black}{#1}}
\newcommand{\CHANGEDNEW}[1]{\textcolor{black}{#1}}
\newcommand{\ASE}[1]{\textcolor{black}{#1}}
\newcommand{\ASEnew}[1]{\textcolor{black}{#1}}
\newcommand{\ASEnn}[1]{\textcolor{black}{#1}}

\definecolor{mygreen}{rgb}{0.0, 0.2, 0.13}
\newcommand{\ASEnnn}[1]{\textcolor{black}{#1}}

\usepackage{marginnote}
\usepackage[backgroundcolor=white,bordercolor=blue,linecolor=blue]{todonotes}
\newcommand{\ASER}[2]{\textcolor{black}{#2}}

\newcommand{\JRR}[2]{\textcolor{black}{#2}}

\newcommand{\TR}[2]{\textcolor{black}{#2}}
\newcommand{\TRC}[1]{\textcolor{black}{#1}}

\newcommand{\IEE}{IEE\xspace}

\newcommand{\GazeDNN}{GazeDNN\xspace}

\newcommand{\CloseDNN}{OC\xspace}

\newcommand{\CropDNN}{CropDNN\xspace}

\newcommand{\GazeDNNR}{GazeDNN\_R\xspace}

\newcommand{\GazeDNNL}{GazeDNN\_L\xspace}

\newcommand{\OC}{OC\xspace}

\newcommand{\EquationsSize}{\small}

\newcommand{\GD}{GD\xspace}
\newcommand{\HPD}{HPD\xspace}
\newcommand{\FLD}{FLD\xspace}

\newcommand{\TrafficDNN}{TSR\xspace}
\newcommand{\ODDNN}{OD\xspace}

\usepackage{url}

\begin{document}

\title{Supporting DNN Safety Analysis and Retraining through Heatmap-based Unsupervised Learning}
%
%
%

\author{Hazem~Fahmy,
        Fabrizio~Pastore,~\IEEEmembership{Member,~IEEE,}
        Mojtaba Bagherzadeh~\IEEEmembership{Member,~IEEE,}
        and~Lionel~Briand,~\IEEEmembership{Fellow,~IEEE}
\thanks{Manuscript received XX YY, 20ZZ; revised MONTH ZZ, 20ZZ.}%
\thanks{H. Fahmy and F. Pastore are affiliated with the SnT Centre for Security, Reliability and Trust, University of Luxembourg, JFK 29, L1855, Luxembourg. (e-mail: hazem.fahmy@uni.lu; fabrizio.pastore@uni.lu).}
\thanks{M. Bagherzadeh is affiliated with the school of EECS, University of Ottawa, Ottawa, ON K1N 6N5, Canada (e-mail: m.bagherzadeh@uottawa.ca).}
\thanks{L. Briand holds shared appointments with the SnT Centre for Security, Reliability and Trust, University of Luxembourg, Luxembourg and the school of EECS, University of Ottawa, Ottawa, ON K1N 6N5, Canada (e-mail: lbriand@uottawa.ca).}}

%
%

\markboth{Accepted for publication on IEEE TRANSACTIONS ON RELIABILITY}%
{Fahmy \MakeLowercase{\textit{et al.}}: Supporting DNN Safety Analysis and Retraining through Heatmap-based Unsupervised Learning}
%



\maketitle

\begin{abstract}
Deep neural networks (DNNs) are increasingly \ASE{important in safety-critical systems,} 
for example in their perception layer to analyze images.
Unfortunately, there is a lack of methods to ensure the functional safety of DNN-based components.

We observe three major challenges with existing practices regarding DNNs in safety-critical systems: 
(1) scenarios that are underrepresented in the test set may lead to serious safety violation risks, but may, however, remain unnoticed; 
(2) characterizing such high-risk scenarios is critical for safety analysis;
(3) retraining DNNs to address these risks is poorly supported when causes of violations are difficult to determine.

To address these problems in the context of DNNs analyzing images, 
we propose \APPR, an approach that automatically
supports the identification of root causes for DNN errors. 
\APPR identifies root causes by 
applying a clustering algorithm to 
heatmaps capturing the relevance of every DNN neuron on the DNN outcome. 
Also, \APPR retrains DNNs with images that are automatically selected based on their relatedness to the identified image clusters.

We evaluated \APPR with DNNs from the automotive domain. \APPR was able to identify all the distinct root causes of DNN errors, thus supporting safety analysis. Also, our retraining approach has shown to be more effective at improving DNN accuracy than existing approaches.
\end{abstract}


%
\IEEEpeerreviewmaketitle

%
%
%
%


\section{Introduction}

\IEEEPARstart{D}{eep} 
Neural Networks (DNNs) are common building blocks in many modern software systems.
This is true for cyber-physical systems, where DNNs are commonly used in their perception layer, and common in the automotive sector, where DNN-based products have shown to effectively automate difficult tasks.
For example, DNNs are used in Advanced Driver Assistance Systems (ADAS) to automate driving tasks such as emergency braking or lane changing~\cite{NVIDIADNN,TeslaDNN}.
\ASEnew{The rise of DNN-based systems concerns manufacturers that produce intelligent car components~\cite{IEE,ZF}. }
This is the case of \IEE~\cite{IEE}, 
\ASEnew{our industry partner in this research}, 
who develops in-vehicle monitoring systems such as drowsiness detection and gaze detection systems~\cite{Naqvi2018}.

DNNs consist of layers of hundreds of neurons transforming high-dimensional vectors through linear and non-linear activation functions, whose parameters are learned during training. 
Such structure prevents engineers from understanding the rationale of predictions through manual inspection of DNNs and, consequently, inhibits software quality assurance practices that rely on the analysis and understanding of the system logic. Such practices include failure root cause analysis and program debugging, which are the target of this paper.


\ASER{3.2}{A root cause is \emph{a source of a defect such that if it is removed, the defect is decreased or removed}~\cite{ISO24765}. 
With DNN-based systems, root cause analysis consists in characterizing system inputs that lead to erroneous DNN results. 
For example, in image classification tasks, a root cause of DNN errors could be severe gender imbalance in the training set leading the DNN to label most female doctors as nurses; 
it might be detected after noticing that error-inducing inputs are characterized by doctors with long hair~\cite{Selvaraju17}.
The DNN can be efficiently retrained after including in the training set additional images featuring these error-inducing characteristics.}


When DNN-based systems are used in a safety-critical context, 
root cause analysis is required to support safety analysis.
Indeed, safety standards, such as ISO26262~\cite{ISO26262} and ISO/PAS 21448~\cite{SOTIF}, enforce the identification of the situations in which the system might be unsafe (i.e., provide erroneous and unsafe outputs) and the design of countermeasures to put in place (e.g., integrating different types of sensors). In the case of DNN-based systems, because of the complex structure of DNNs, the clear identification of unsafe situations is a challenge.

When inputs are images, which is our focus here, existing solutions for root cause analysis generate heatmaps that use colors to capture the importance of pixels in their contribution to a DNN result~\cite{Selvaraju17,Montavon2019}. 
By inspecting the heatmaps generated for a set of erroneous results, a human operator can determine that these heatmaps highlight the same objects, which may suggest the root cause of the problem (e.g., long hair~\cite{Selvaraju17}). 
Based on the identified root cause, engineers can then retrain the DNN using additional images with similar characteristics. 
Unfortunately, this process is expensive and error-prone because it relies on the visual inspection of many generated heatmaps. 
MODE goes beyond visual inspection and supports the automated debugging of DNNs through the identification of likely error-inducing images to be used for retraining~\cite{Ma2018}. 
However, MODE cannot support safety analysis since it does not provide support to identify plausible and distinct root causes leading to DNN errors.

To alleviate the limitations above, we propose Heatmap-based Unsupervised Debugging of DNNs (\APPR). 
\ASER{3.2a}{\APPR relies on hierarchical agglomerative clustering~\cite{King:2014} combined with a specific heatmap-based distance function to identify clusters of error-inducing images with similar heatmaps for internal layers.
Since heatmaps capture the importance of neurons regarding their contribution to the DNN result, error-inducing images with similar heatmaps should share characteristics that drive the generation of erroneous DNN results.}
\ASE{Each cluster} should thus characterize a distinct root cause for the observed DNN errors, even in cases where such causes are infrequent. Images in such clusters should then help identify clear and distinct root causes and can serve as a basis for efficient and effective retraining. 
\ASE{We focus on internal DNN layers because they act as an abstraction over the inputs (e.g., ignore image background).}


More precisely, \APPR relies on the computed clusters to identify new images to be used to retrain the DNN. 
\ASE{ Given a potentially large set of collected or generated unlabeled images, \APPR 
selects the subset of images that are closer to the identified clusters according to a heatmap-based distance.
These images are then labeled by engineers and used to retrain the network. Labeling only a subset of images reduces retraining cost.
}


We performed an empirical evaluation on {six} DNNs.
Our empirical results show that \APPR can automatically and accurately identify the different root causes of DNN errors.
Also, our results suggest that the \APPR retraining process, improves DNN accuracy up to {30.24} percentage points and is more effective than baseline approaches.

The paper is structured as follows. Section~\ref{sec:context} provides the context and motivation for this work. Section~\ref{sec:background} summarizes background information. 
\ASE{Section~\ref{sec:approach} presents the proposed approach in details.}
Section~\ref{sec:empirical} reports on the results of our empirical evaluation. Section~\ref{sec:related} discusses related work. Section~\ref{sec:conclusion} concludes the paper.

\section{Motivation and Context}
\label{sec:context}

In this section, we introduce the practical context of our research, which in short is the safety analysis and debugging of DNN-based automotive systems.
We explain why automated root cause analysis is necessary to enable functional safety analysis. 
Also, we show how DNN accuracy improvement can be facilitated by the automated characterization and identification of \emph{error-inducing inputs} (i.e.,
inputs that make the DNN generate erroneous results).
Though the issues raised below and many of our insights are not specific to automotive systems, but also relevant to many cyber-physical systems in general, this is the practical domain and context in which this work took place.

\begin{figure}[htbp]
\begin{minipage}{3cm}
\includegraphics[width=3cm]{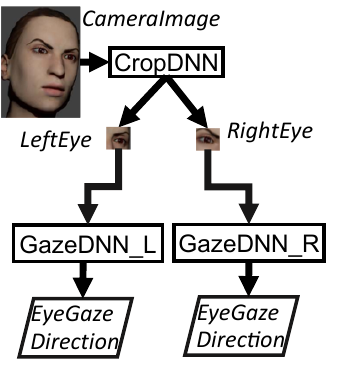}
\caption{DNN-based system for gaze detection.}
\label{fig:DNNs}
\end{minipage}
\hspace{2mm}
\begin{minipage}{5cm}
\includegraphics[width=5cm]{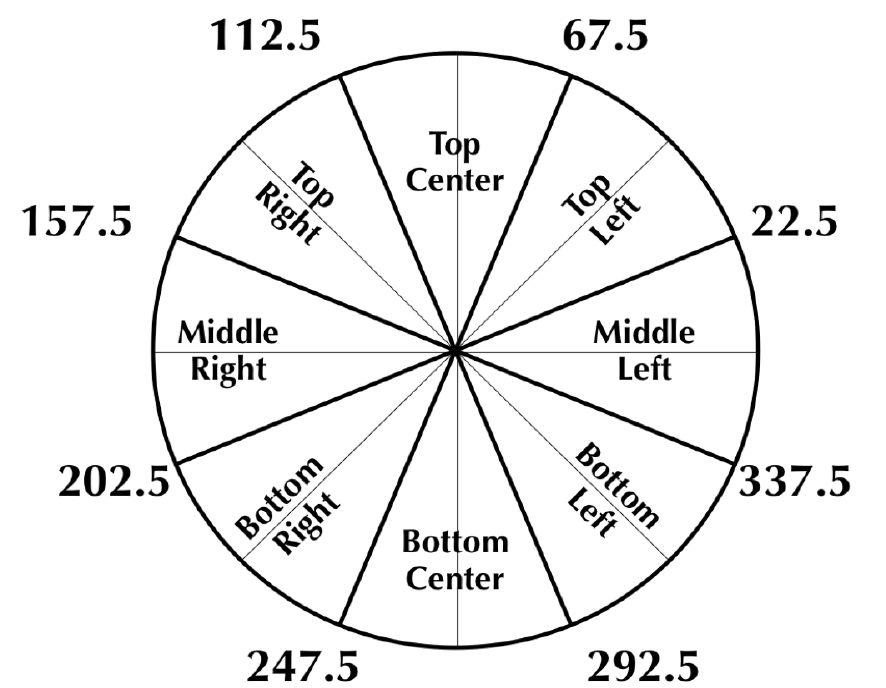}
\caption{Gaze directions.}
\label{fig:angles}
\end{minipage}
\end{figure}

\subsection{DNN-based automotive systems}
\label{sec:context:case}

Our work is motivated by the challenges encountered in industry sectors developing safety-critical, cyber-physical systems, such as the automotive sector. For example, this is the case for \IEE \cite{IEE}, a supplier of sensing solutions active in the automotive market and the provider of our case studies.
In particular, \IEE develops a gaze detection system (\GDS) which uses DNNs to determine the gaze direction of the driver, from images captured by a camera on the instrument panel of the car.

\IEE has been evaluating the feasibility of different \GDS system architectures. 
Figure~\ref{fig:DNNs} shows an architecture consisting of three DNNs (i.e., \CropDNN, \GazeDNNL, \GazeDNNR). 
\CropDNN identifies face landmarks that enable the cropping of images containing the eyes only. 
\GazeDNNL and \GazeDNNR classify the gaze direction into eight classes (i.e., 
TopLeft, TopCenter, TopRight, MiddleLeft, MiddleRight, BottomLeft, BottomCenter, and BottomRight). 


To reduce training costs, \IEE
relies on training sets containing images that are collected from driving scenes and images generated by simulation software.
Simulators are used to reduce the costs related to data collection and data labeling. Indeed, models of the dynamics of real-world elements (e.g., eyeballs) are used to generate, in a controlled way through the selection of parameter values, hundreds of images in a few hours~\cite{Wood2016}. 
Further, and this is important in terms of cost saving, simulation enables the automated labeling of images by analyzing model parameters.
\ASEnew{However, while simulator images alleviate the costs of training, testing ultimately requires real-world images as well since simulators do not exhibit perfect fidelity.}

In our experiments with \IEE, we rely on the UnityEyes simulator to generate eye images~\cite{Wood2016}. 
UnityEyes combines a generative 3D model of the human eye region with a real-time rendering framework \CHANGEDNEW{based on Blender~\cite{Blender}.}
We determine the gaze direction label from the gaze angle parameter provided by UnityEyes, based on predefined gaze ranges depicted in Figure~\ref{fig:angles}.
For example, we assign the label \emph{TopCenter} when the gaze angle is between 67.5 and 112.5 degrees.

\CHANGEDNEW{Additional DNN-based systems under development at IEE, which we used as cases studies, are presented in Section~\ref{sec:empirical}.}


%

\subsection{Debugging of DNN-based Systems}
\label{sec:context:debugging}

\begin{figure}[tb]
\includegraphics[width=8.4cm]{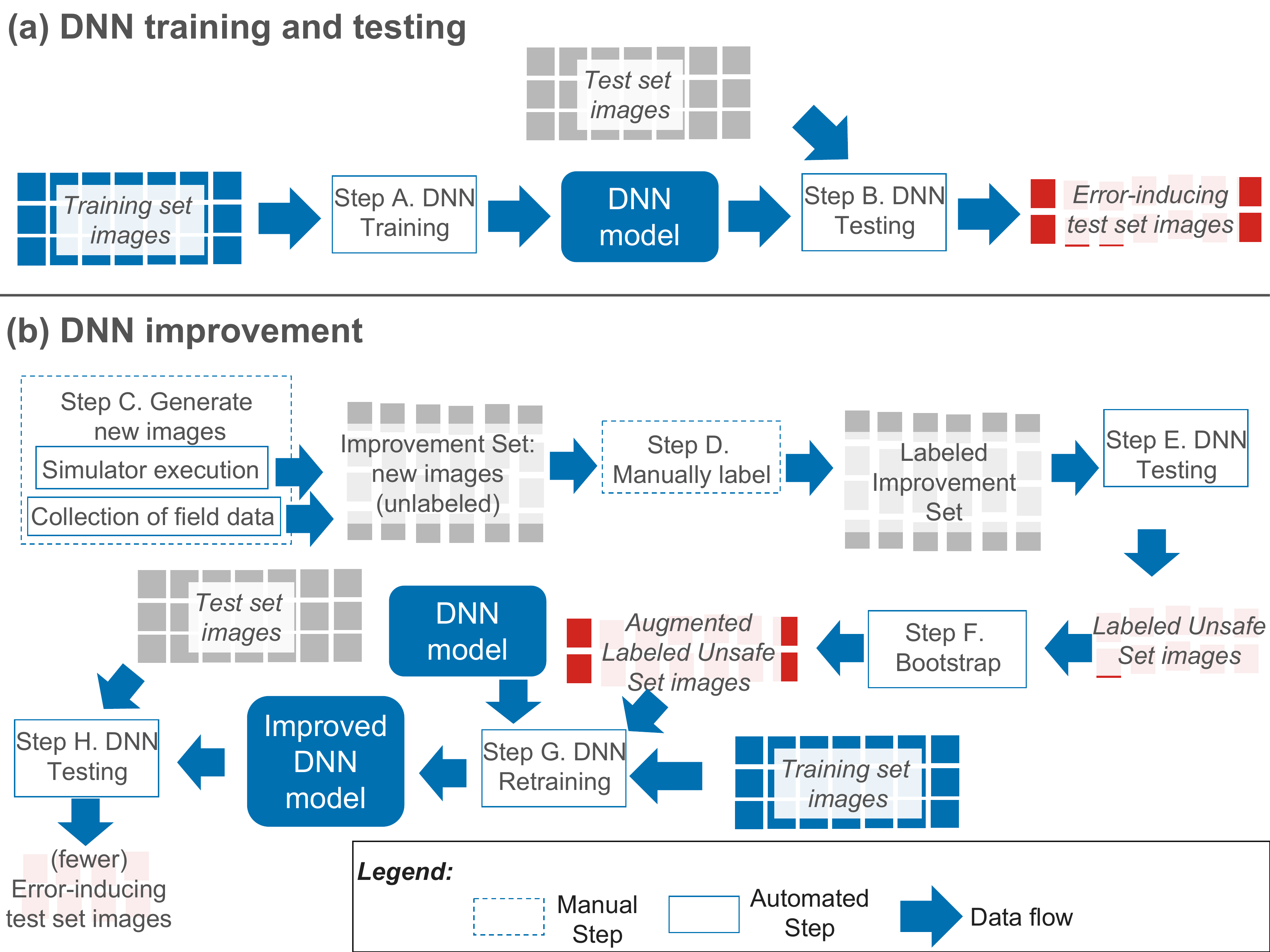}
\caption{Training and debugging DNNs.}
\label{fig:TraditionalApproach}
\end{figure}

\IEE engineers train the DNNs that compose their systems by following the standard machine learning process depicted in Figure~\ref{fig:TraditionalApproach}-a.
They first train the DNN using a training set with labeled images (Step A) and then execute the DNN against a labeled test set (Step B). 
This process enables engineers to evaluate the DNN accuracy (e.g., the percentage of images leading to correct results).

When the accuracy of the system is not adequate, engineers typically improve the DNN by augmenting the training set with error-inducing images.
This process is depicted in Figure~\ref{fig:TraditionalApproach}-b.
First, engineers generate a set of new images to be used to retrain the DNN (Step C). We call this set of images \emph{improvement set}.
The improvement set generally consists of images collected from the field since these tend to be error-inducing when DNNs have been trained using simulator images.
Real-world images must be manually labelled (Step D). 
The DNN model is tested with the improvement set and images that lead to DNN errors are identified (Step E). 
This set of error-inducing (unsafe) images is considered to retrain the DNN (Step G), 
using as initial configuration for DNN weights the ones in the previously trained model. 

To improve the DNN, it is necessary to process a sufficiently large number of unsafe images.
For this reason, the number of unsafe images can be augmented by applying \emph{bootstrap resampling} (i.e., by replicating samples in the unsafe set~\cite{ML:Book}) till a target is achieved (Step F).
Finally, the improved DNN can be assessed on the test set (Step H).

In general, generating a sufficiently diverse set of error-inducing inputs that include all possible causes of DNN errors is very difficult. 
Further, when the labeling of such images is manual, the costs of labeling becomes prohibitive and DNN improvement is hampered.
For this reason, \textbf{automatically characterizing images that are likely to lead to DNN errors} would allow the image generation or selection process to target specific types of images and increase the efficiency of the DNN retraining process.

%
%

\subsection{Functional safety analysis}
\label{sec:context:fsa}
%
%
%
%


Like many other organizations in the automotive domain, \IEE products must comply with the functional safety standards ISO 26262 and ISO/PAS 21448.
Functional safety is addressed by identifying, for each component of the product (e.g., the DNNs of the \GDS), the \emph{unsafe conditions} 
that could lead to hazards, by identifying countermeasures (e.g., redundant components), and by demonstrating that these unsafe conditions are unlikely to occur.

ISO/PAS 21448, specifically targeting autonomous systems, recommends to determine unsafe conditions by following the traditional DNN testing process depicted in
Figure~\ref{fig:TraditionalApproach}-a and by manually inspecting the error-inducing images to look for root causes of DNN errors.
In a DNN context, \emph{unsafe conditions thus correspond to root causes of DNN errors}.

According to ISO/PAS 21448, engineers can set a quantitative target for accuracy evaluation to demonstrate that unsafe situations are unlikely. 
However, ISO/PAS 21448 also points out that quantitative targets are not sufficient and that engineers remain liable for potentially hazardous scenarios missing from the test set.


In addition, the manual identification of unsafe conditions is error-prone. For example, engineers may overlook unsafe conditions that are underrepresented in the test set. Also, such conditions may lead to a biased estimate of the accuracy of the DNN.
For example, UnityEyes generates eye images where the horizontal angle of the head is determined based on a uniform distribution, between 160 (head turned right) and 220 degrees (head turned left). 
As a result, very few images with an angle of 160 or 220 degrees are generated and, though it may be an unsafe condition (i.e., one eye is barely visible and the gaze direction prediction may be inaccurate),
experiments based on test sets generated with UnityEyes may suggest that the DNN is on average very accurate. It is, however, important for engineers to know that such a DNN, in some rarely occurring cases in the test set, is unsafe when the driver turns his head while driving.

In summary, accuracy estimation results depend on the test set, which may not include all unsafe conditions in a representative or balanced manner. 
Automated root cause analysis helps making sure, through clustering, that even rare, unsafe conditions are made visible to the analyst, especially when safety analysis time is limited. In other words, clustering based on heatmaps makes safety analysis robust, to some extent, to imperfect test sets.




\section{Background}
\label{sec:background}

\subsection{DNN Explanation and Heatmaps}
 \label{sec:background:explanation}
Approaches that aim to explain DNN results have been developed in recent years~\cite{GARCIA2018}. 
Most of these concern the generation of heatmaps that capture the importance of pixels in image predictions. They include black-box~\cite{Petsiuk2018rise,Dabkowski17} and white-box
approaches~\cite{Montavon2019,Selvaraju17,Zeiler14,DB15a,Zhou16}. 
Black-box approaches generate heatmaps for the input layer and do not provide insights regarding internal DNN layers.
In this paper, we therefore resort to white box approaches which rely on the backpropagation of the relevance score computed by the DNN~\cite{Montavon2019,Selvaraju17,Zeiler14,DB15a,Zhou16}; 
Castanon et al. provide an overview of the state of the art ~\cite{Castanon18}. 
In this paper, we rely on Layer-Wise Relevance Propagation (LRP)~\cite{Montavon2019} because of the limitations of other approaches.
First, solutions ~\cite{Zhou16} backpropagating only the difference in activations between the different classes may compromise clustering \CHANGEDNEW{since they do not account for information about all available neurons but only the ones related to the predicted output class}. 
Deconvolutional networks~\cite{Zeiler14} and guided backpropagation~\cite{DB15a} 
lead to sparse heatmaps that do not fully explain the DNN result~\cite{Samek17}.
Grad-CAM~\cite{Selvaraju17} does not work with convolutional DNN layers.
In contrast, LRP generates precise, non-sparse heatmaps for all the DNN layers because
it takes into account all the different factors affecting the relevance of a neuron, which include the DNN structure and the neuron activations.


\begin{figure}[b]
\includegraphics[width=8.4cm]{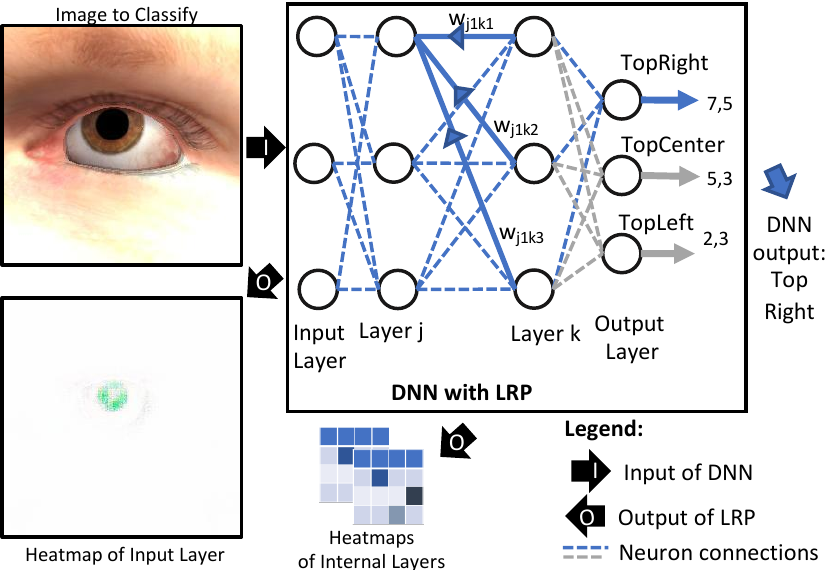}
\caption{Layer-Wise Relevance Propagation.}
\label{fig:LRP}
\end{figure}

LRP redistributes the relevance scores of neurons in a higher layer to those of the lower layer. Assuming $j$ and $k$ to be two consecutive layers of the DNN, LRP propagates the relevance scores computed for a given layer $k$ into a neuron of the lower layer $j$. It has been theoretically justified as a form of Taylor decomposition~\cite{MONTAVON2017DTD}.

Figure~\ref{fig:LRP} illustrates the execution of LRP on a fully connected network used to classify inputs. 
LRP analyzes the data processed by a DNN and can be applied to any DNN architecture.
In the forward pass, the DNN receives an input and generates an output (e.g., classifies the gaze direction as TopLeft) while keeping trace of the activations of each neuron.
The heatmap is generated in a backward pass. 

In Figure~\ref{fig:LRP},
blue lines show that the DNN score of the selected class is backpropagated to lower layers. Plain lines show the connections concerned by the propagation formula used to compute the relevance ($R_{ji}$) of neuron $i$ at layer $j$ from all the connected neurons in layer $k$.
$R_{ji} = \sum_{l}(\frac{z_{ji\_kl}}{\sum_{i}z_{ji\_kl}}*R_{kl})$, where $z_{ji\_kl}$ captures the extent to which neuron $ji$ has contributed to make neuron $kl$ relevant, and $R_{kl}$ captures the relevance of neuron $l$ at layer $k$.
\CHANGED{For example, for linear layers, $z_{ji\_kl} = a_{ji}*w_{ji\_kl}^{+}$, where $a_{ji}$ is the activation of neuron $i$ at layer $j$ and $w_{ji\_kl}^{+}$ is the value of the weight on the connection between neuron $ji$ and neuron $ki$, considering positive weights only. 
The denominator is used to redistribute the relevance received by a neuron to the lower layer proportionally to relative contributions.}
\CHANGED{In our experiments, we have applied LRP to Convolutional Neural Networks (CNNs~\cite{Goodfellow2016}), for classification tasks, and Hourglass Neural Networks~\cite{Hourglass}, for regression tasks.
 We rely on the LRP and $z_{ji\_kl}$ implementation provided by LRP authors~\cite{WIFS2017}.}

The heatmap in Figure~\ref{fig:LRP}  shows that the result computed by the DNN was mostly influenced by the pupil and part of the eyelid, which are the non-white parts in the heatmap.

An additional key benefit of LRP is that it enables the computation of \emph{internal heatmaps}, i.e., heatmaps for the internal layers of the DNN,
based on the relevance score computed for every neuron in every layer. 
An internal heatmap for a layer $k$ consists of a matrix with the relevance scores computed for all the neurons of layer $k$.

\subsection{Unsupervised Learning}
\label{sec:back:cluster}
\ASE{Unsupervised learning concerns the automated identification of patterns in data sets without pre-existing labels.}
In this paper, we rely on hierarchical agglomerative clustering (HAC)~\cite{King:2014} to identify groups of error-inducing images with similar characteristics.

HAC is a bottom up approach in which each observation starts in its own cluster and pairs of clusters are iteratively merged into a sequence of nested partitions. 
The input of HAC is a matrix capturing the distance between every observation pair.
The grouping that occurs at each step aims to minimize an objective function.
In HAC, widely adopted objective functions are (1) the error sum of squares within clusters (i.e.,Ward’s linkage method~\cite{Ward}), to help minimize within-cluster variance, (2) the average of distances between all pairs of elements belonging to distinct clusters (i.e., average linkage~\cite{UPGMA}), to help maximize diversity among clusters, and (3) the shortest distance between a pair of elements in two clusters (i.e., single linkage~\cite{ClusteringAlgos}), to merge clusters that are closer for at least one element.

HAC leads to a hierarchy of clusters that can be represented as a dendrogram. 
\CHANGED{To identify the \emph{optimal number of clusters}, 
we rely on the knee-point method~\cite{SatopaaKNEE11}, a recent approach that has been applied in different contexts, including 
fault localization~\cite{LiRootCause19} and performance optimization~\cite{JendeleDecomposition19}.
It automates the \emph{elbow method} heuristics~\cite{Thorndike1953}, which is commonly used in cluster analysis and consists of plotting the variance within a cluster as a function of the number of clusters, and picking the curve's elbow as the number of clusters to use.
The knee-point method automates it by fitting a spline to raw data through univariate interpolation and normalizing min/max values of the fitted data.
The knee-points are the points at which the curve differs most from the straight line segment connecting the first and last data point.}



\TR{2.1}{We chose HAC over K-means~\cite{mcqueen1967smc} since 
the latter requires the number of clusters to be known or predicted. In our case, K-means would thus need to be repeatedly executed in order to determine the optimal number of clusters; in the case of HAC, instead, the generated dendrogram provides all the required information in a single run.
Further, HAC does not require the computation of cluster centroids~\cite{Murtagh2014}, which is particularly expensive when differences between observations are computed from large matrices~\cite{manning2008introduction}.
To more formally motivate our choice, we compare the worst case time complexity of HAC and K-means. HAC's running time is in $\BigO \Big( \frac{d \cdot n\cdot(n-1)}{2}+n^2\Big) \approx \BigO \Big( d \cdot n^2\Big)$, with $n$ being the number of instances to cluster, and $d$ being number of features to consider during clustering (i.e., the entries of the heatmap matrix, see Section~\ref{sec:appr:clustering}). In other words, this complexity depends on the cost of computing the distance matrix, which is equal to the number of features multiplied by the number of image pairs (first addend), and the time complexity of HAC with Ward linkage (second addend~\cite{Murtagh2014}).
The worst case time complexity of a single iteration for K-means is $\BigO \Big( n^{k \cdot d} \Big)$, with $k$ being the number of clusters to consider~\cite{Inaba2000VariancebasedKA,Vattani2011}. 
In our experiments, $n$ lies in the range [506-5371], the \emph{optimal number of clusters} is between 11 and 20 (see Section~\ref{sec:evaluation:RQ1.1}), and $d$ is very large for DNN convolutional layers (e.g., $169 \times 256$, see Section~\ref{sec:appr:clustering}). These numbers show that the worst case complexity of K-means is much larger than that of HAC, which further motivates our choice. We leave to future work the empirical evaluation of K-means and other clustering solutions.}

\section{The \APPR Approach}
\label{sec:approach}

\begin{figure}[b]
\includegraphics[width=8.4cm]{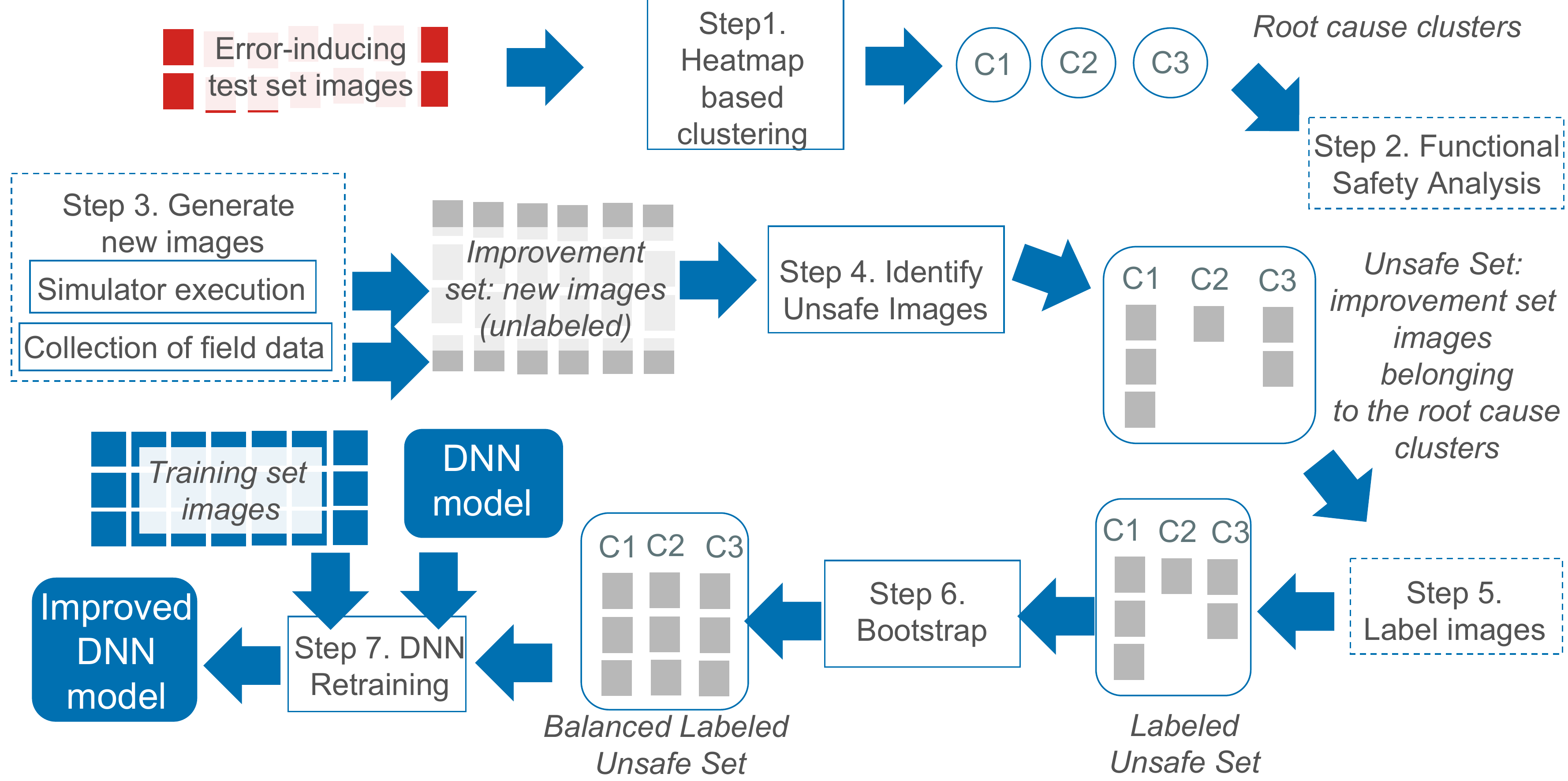}
\caption{Overview of \APPR (legend in Fig.~\ref{fig:TraditionalApproach}).}
\label{fig:Approach}
\end{figure}

Figure~\ref{fig:Approach} provides an overview of our approach, \APPR, 
\JRR{R1.2}{which provides two main contributions: (1) it automatically identifies the root causes of DNN errors, (2) it automatically identifies unsafe images for retraining the DNN.
\APPR consists of six steps, described below.}

In Step 1, \APPR performs heatmap-based clustering. This is a core contribution of this paper and consists of three activities: (1) generate heatmaps for the error-inducing test set images, (2) compute distances between every pair of images using a distance function based on their heatmaps, and (3) execute hierarchical agglomerative clustering to group images based on the computed distances. Step 1 leads to the identification of root cause clusters, i.e., clusters of images with a common root cause for the observed DNN errors.

In Step 2, engineers inspect the root cause clusters (typically a small number of representative images) to identify unsafe conditions, as required by functional safety analysis. \JRR{R1.1}{The inspection of root cause clusters is an activity performed to gain a better understanding of the limitations of the DNN and thus introduce countermeasures for safety purposes (see Section~\ref{sec:context:fsa}), if needed. However, the inspection of root cause clusters has no bearing on the later steps of our approach, including retraining.}

In Step 3, engineers rely on real-world data or simulation software
\ASE{to generate a new set of images to retrain the DNN, referred to as the \emph{improvement set}.}

\ASE{In Step 4, \APPR \emph{automatically} identifies the subset of images belonging to the improvement set that are likely to lead to DNN errors, referred to as the  \emph{unsafe set}. It is obtained by assigning 
the images of the improvement set to the root cause clusters according to their heatmap-based distance.} 

In Step 5, engineers manually label the images belonging to the unsafe set, if needed (e.g., in the case of real images). Different from traditional practice (see Figure~\ref{fig:TraditionalApproach}-b), \APPR requires that engineers label only a small subset of the improvement set. 

In Step 6, to improve the accuracy of the DNN for every root cause observed, 
regardless of their frequency of occurrence in the training set,
\APPR balances the labeled unsafe set using a bootstrap resampling approach.

In Step 7, the DNN model is retrained by relying on a training set that consists of the union of the original training set and the balanced labeled unsafe set.



\TR{1.2}{Three out of the seven steps are manual (Steps 2, 3, and 5). However, these steps are also part of state-of-the-art solutions (see Section~\ref{sec:context:debugging}). But in the case of HUDD the manual effort required in such steps is much more limited than in existing approaches. 
With \APPR, in Step 2, engineers inspect a few images per root cause clusters rather than the whole set of images, thus resulting in (a) significant cost savings (see Section~\ref{sec:evaluation:RQ1.1}) and (b) effective guidance towards the identification of root causes. 
In Step 5, with \APPR, engineers label only a subset of the improvement set that contains likely unsafe images identified by \APPR. Such unsafe images can be effectively used for retraining. Without \APPR, engineers would label a randomly selected subset of the improvements set, which would likely contain less unsafe images and thus be less effective during retraining (see Section~\ref{fig:Baseline2}).
Finally, Step 3 is common practice and entails limited effort (e.g., buying field images or configuring a simulator).}

\JRR{R1.2}{The quality of HUDD results does not depend on the personal ability of engineers involved in manual steps; indeed, manual steps either concern the inspection of HUDD results or involve simple activities. The first contribution of HUDD (i.e., identify root causes of DNN errors) is provided by Step 1, which is fully automated. Step 2, which is manual, concerns the visual inspection of the generated clusters, does not require particular skills, and is part of state-of-the-art approaches. However, with HUDD, this step is facilitated by the quality of the generated clusters; for example, in Section~\ref{sec:emp:clusters} we demonstrate that HUDD generates root cause clusters presenting a common set of characteristics that are plausible causes of DNN errors, thus facilitating the identification of these root causes. The other manual steps (i.e., Step 3 and Step 5) are simple. In Step 3, engineers simply generate additional images (their selection is automated by HUDD in Step 4). In Step 5, engineers provide additional labels, which is an activity that, despite being time-consuming, can be assumed to be correct most of the time and is unavoidable when supervised learning (e.g., DNNs) is involved. The other steps leading to the second contribution of HUDD (i.e., identification of unsafe images and DNN retraining), are fully automated.}

The following sections describe in detail all the steps of the approach, except Steps 3 and 5, which were introduced in Section~\ref{sec:context:debugging}.


\subsection{Heatmap-based clustering}
\label{sec:appr:clustering}

\begin{figure}[b]
\includegraphics[width=8.4cm]{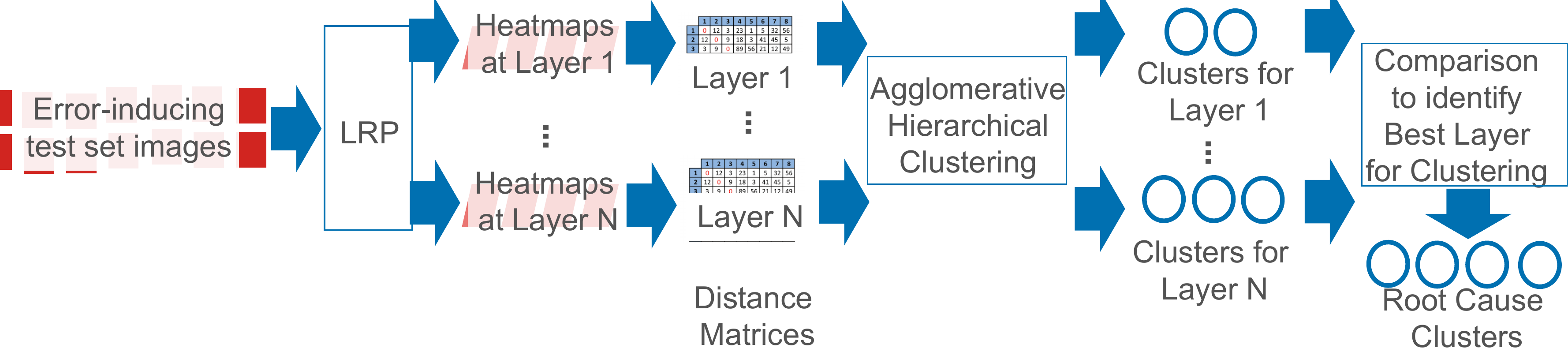}
\caption{Heatmap-based Clustering}
\label{fig:Clustering}
\end{figure}

\APPR is based on the intuition that, since heatmaps capture the  relevance of each neuron on DNN results, error-inducing inputs sharing 
the same root cause should show similar heatmaps.
For this reason, to identify the root causes of DNN errors, we rely on clustering based on heatmaps. 
Figure~\ref{fig:Clustering} provides an overview of our clustering approach.

\TR{1.1}{For each error-inducing image in the test set, \APPR relies on LRP to generate heatmaps of internal DNN layers. Each heatmap captures the relevance score of each neuron in that layer.} 

\TRC{A heatmap is a matrix with entries in $\mathbb{R}$, i.e., it is a triple $(N,M,f)$ where $N,M \in \mathbb{N}$ and $f$ is a map $[N] \times [M] \rightarrow \mathbb{R}$. 
We use the syntax $H[i,j]_x^L$ to refer 
 to an entry in row $i$ (i.e., $i < N$) and column j (i.e., $j < M$) of a heatmap $H$ computed on layer $L$ from an image $x$. 
 The size of the heatmap matrix (i.e., the number of entries) is $N \cdot M$, with $N$ and $M$ depending on the dimensions of the DNN layer L. For convolution layers, $N$ captures the number of neurons in the feature map, while $M$ captures the number of feature maps. For example, the heatmap for the eighth layer of AlexNet has size $169 \times 256$ (convolution layer), while the the heatmap for the tenth layer has size $4096 \times 1$ (linear layer).}

\TRC{Since distinct DNN layers lead to entries defined on different value ranges~\cite{MONTAVON2017DTD}, to enable the comparison of clustering results across different layers, we generate normalized heatmaps by relying on min-max normalization~\cite{DataMiningBook}. For a layer L, we identify $\mathit{min}_L$ as the minimum value observed for all the heatmaps generated for a layer $L$, i.e.,  
\begin{equation}
\EquationsSize
\mathit{min}_L \le H[i,j]_x^L \ \bigforall \ i < N, j < M
\end{equation}
with $x$ being an image belonging to the unsafe test set. The maximum value $\mathit{max}_L$ is derived accordingly. An entry $\tilde{H}[i,j]_x^L$ belonging to a normalized heatmap $\tilde{H}_x^T$ is derived as 
\begin{equation}
\EquationsSize
\tilde{H}[i,j]_x^L = \frac{H[i,j]_x^L-min_L}{\mathit{max}_L-\mathit{min}_L}
\end{equation}
The generated normalized heatmaps are used to build, for each DNN layer, a distance matrix  that captures the distance between every pair of error-inducing image in the test set. 
The distance between a pair of images $\langle a,b \rangle$, at layer $L$, is computed as follows:
\begin{equation}
\EquationsSize
\mathit{heatmapDistance}_L(a,b)=\mathit{EuclideanDistance}(\tilde{H}^L_a,\tilde{H}^L_b)
\end{equation}
where $\tilde{H}^L_x$ is the normalized heatmap computed for image $x$ at layer $L$.} 

$\mathit{EuclideanDistance}$ is a function that computes the euclidean distance between two $N \times M$ matrices according to the formula 
\begin{equation}
\EquationsSize
\mathit{EuclideanDistance}(A,B)=\sqrt{ \sum_{i=1}^{N} \sum_{j=1}^{M}  (A_{i,j} - B_{i,j})^{2} }
\end{equation}
where $A_{i,j}$ and $B_{i,j}$ are the values in the cell at row $i$ and column $j$ of the matrix.


\CHANGED{Since we aim to generate clusters including images with similar characteristics, for each layer, we identify clusters of images by relying on the HAC algorithm \ASEnew{with Ward linkage}, which minimises within cluster variance (see Section~\ref{sec:back:cluster}).
We select the optimal number of clusters for a layer using the \emph{knee-point method} applied to the weighted average intra-cluster distance.}


In our context, clustering results are informative if they group together images that are misclassified for a same reason (i.e., if clusters are cohese) and if similar images belong to a same cluster (i.e., clusters are not fragmented).
We determine cluster cohesion based on the weighted average intra-cluster distance ($\mathit{WICD}$), which we define according to the following formula:
\begin{equation}
\EquationsSize
\label{eq:AvgICD}
\mathit{WICD}(L_l)=\frac{\sum^{|L_l|}_{j=1}\bigg( ICD(L_l,C_j)*\frac{|C_j|}{|C|} \bigg) }{|L_l|} 
\end{equation}
where $L_l$ is a specific layer of the DNN, $|L_l|$ is the number of clusters in the layer $L_l$, $ICD$ is the intra-cluster distance for cluster $C_i$ belonging to layer $L_l$, 
$|C_j|$  is the number of elements in cluster $C_j$, while $|C|$ is the number of images in all the clusters.

In Formula~\ref{eq:AvgICD}, $\mathit{ICD}(L_l,C_j)$ is computed as follows:
\begin{equation}
\EquationsSize
\label{eq:ICD}
\mathit{ICD}(L_l,C_j)=\frac{\sum^{N_j}_{i=0}\mathit{heatmapDistance}_{L_{l}}(p^a_i,p^b_i)}{N_j}
\end{equation}
where $p_i$ is a unique pair of images in cluster $C_j$, and $N_j$ is the total number of pairs it contains. The superscripts $a$ and $b$ refer to the two images of the pair to which the distance formula is applied.

\CHANGED{In Formula~\ref{eq:AvgICD}, the factor $\frac{|C_j|}{|C|}$ normalizes the average ICD with respect to the relative size of the cluster. It helps determine the optimal number of clusters within a layer and enables the identification of the best clustering result across layers, as explained in the following paragraphs.}

\begin{figure}
\subfloat[ICD]{\includegraphics[width=.5\linewidth]{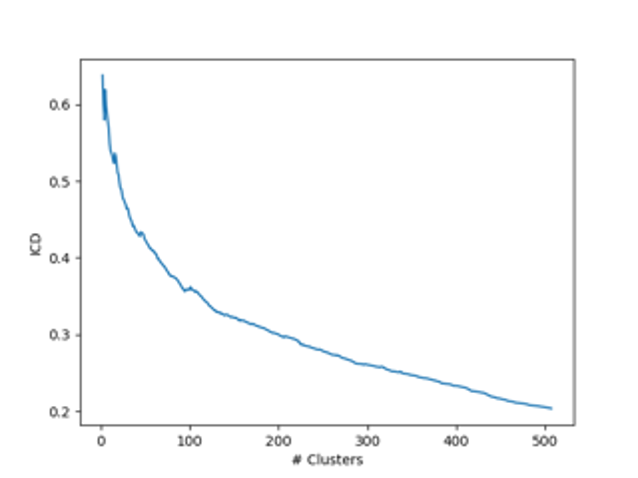}}
\subfloat[WICD]{\includegraphics[width=.5\linewidth]{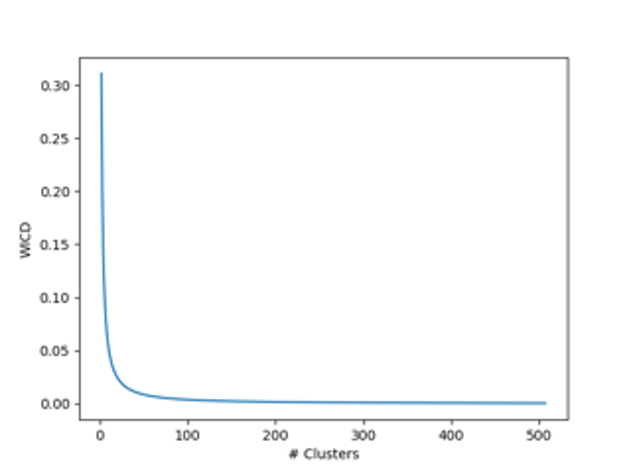}}
\caption{Slope of ICD and WICD for GazeDNN}
\label{fig:appr:clustering}
\end{figure}

\CHANGED{Since Ward linkage groups together elements that minimize within-cluster variance, an increase in the number of clusters (i.e., less elements per cluster) leads to a proportional decrease in the average ICD. In other words, the ICD slope is mild and smooth, which complicates the identification of the optimal number of clusters through the elbow method. 
By taking into account the relative size of the cluster, WICD helps determine when a larger number of clusters leads to suboptimal results, which happens when an increased number of cluster does not break down large clusters but rather divide small clusters into tiny ones (i.e., they are fragmented).
Figure~\ref{fig:appr:clustering} shows the slope obtained for both ICD and WICD for a growing number of clusters; the plot for WICD clearly helps identify the sub-range on the X-axis leading to a drastic change in the slope, thus enabling the identification of an optimal number of clusters beyond which WICD barely decreases.} 

\begin{figure}
\centering
\includegraphics[width=.5\linewidth]{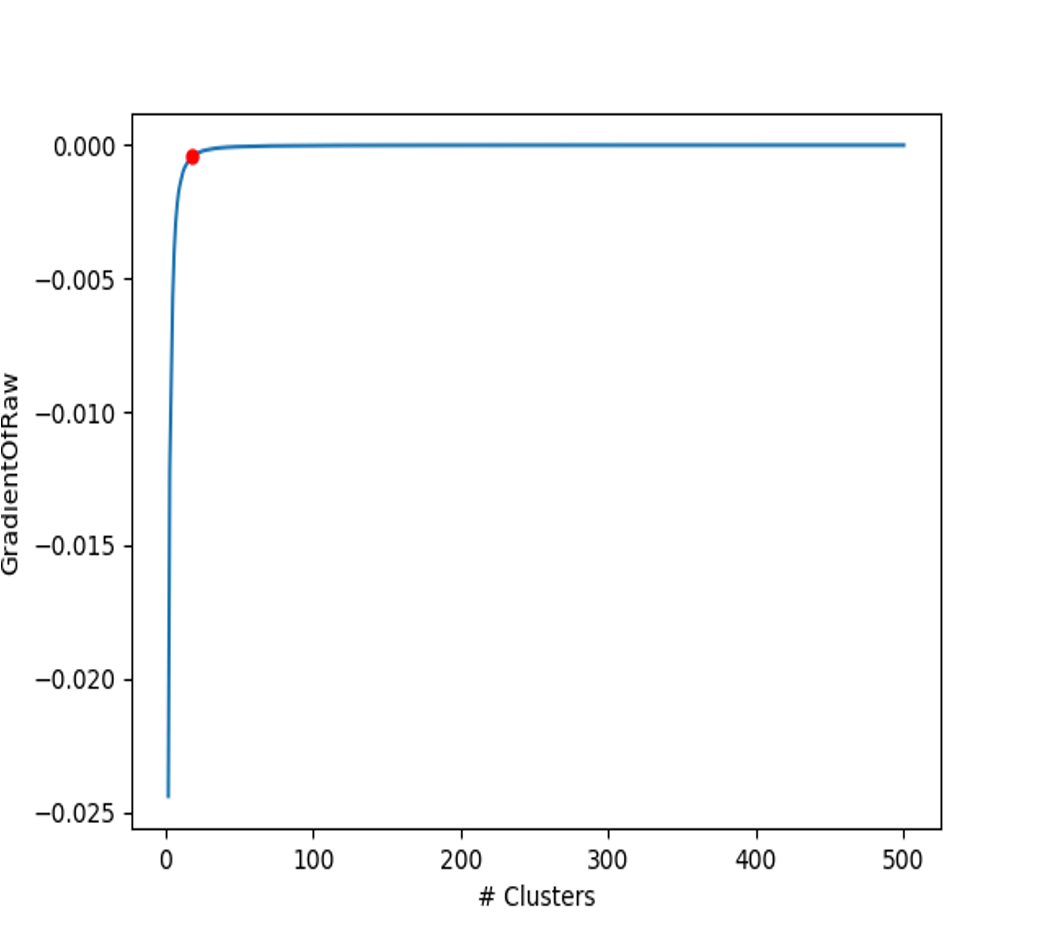}
\caption{Gradient and knee-point (in red) for Figure~\ref{fig:appr:clustering}-b}
\label{fig:appr:clustering:knee}
\end{figure}

\CHANGED{To determine when WICD stops decreasing significantly, we rely on its derivative that we approximate by relying on the fourth order central difference method~\cite{Fornberg88}. We then rely on the knee-point method to identify the point with the maximum curvature in the derivative~\cite{SatopaaKNEE11}.
Figure~\ref{fig:appr:clustering:knee} shows an example knee-point automatically identified with our method.}

\APPR selects 
the layer $L_m$ with the minimal $\mathit{WICD}$. 
By definition, the clusters generated for layer $L_m$ are the ones that maximize cohesion and we therefore expect them to group together images that present similar characteristics, suggesting root causes for DNN errors.

When comparing clusters for distinct layers, the normalization based on the relative size of the cluster (i.e.,  the factor $\frac{|C_j|}{|C|}$ in Equation~\ref{eq:AvgICD})
 enables \APPR to penalize layers including large clusters with high $\mathit{ICD}$. These clusters group together images with heatmaps that are different from each other and thus may be associated with different root causes for DNN errors.


\subsection{Root Causes Inspection}
\label{sec:rootCausesInspection}

\begin{figure}[tb]
\includegraphics[width=8.4cm]{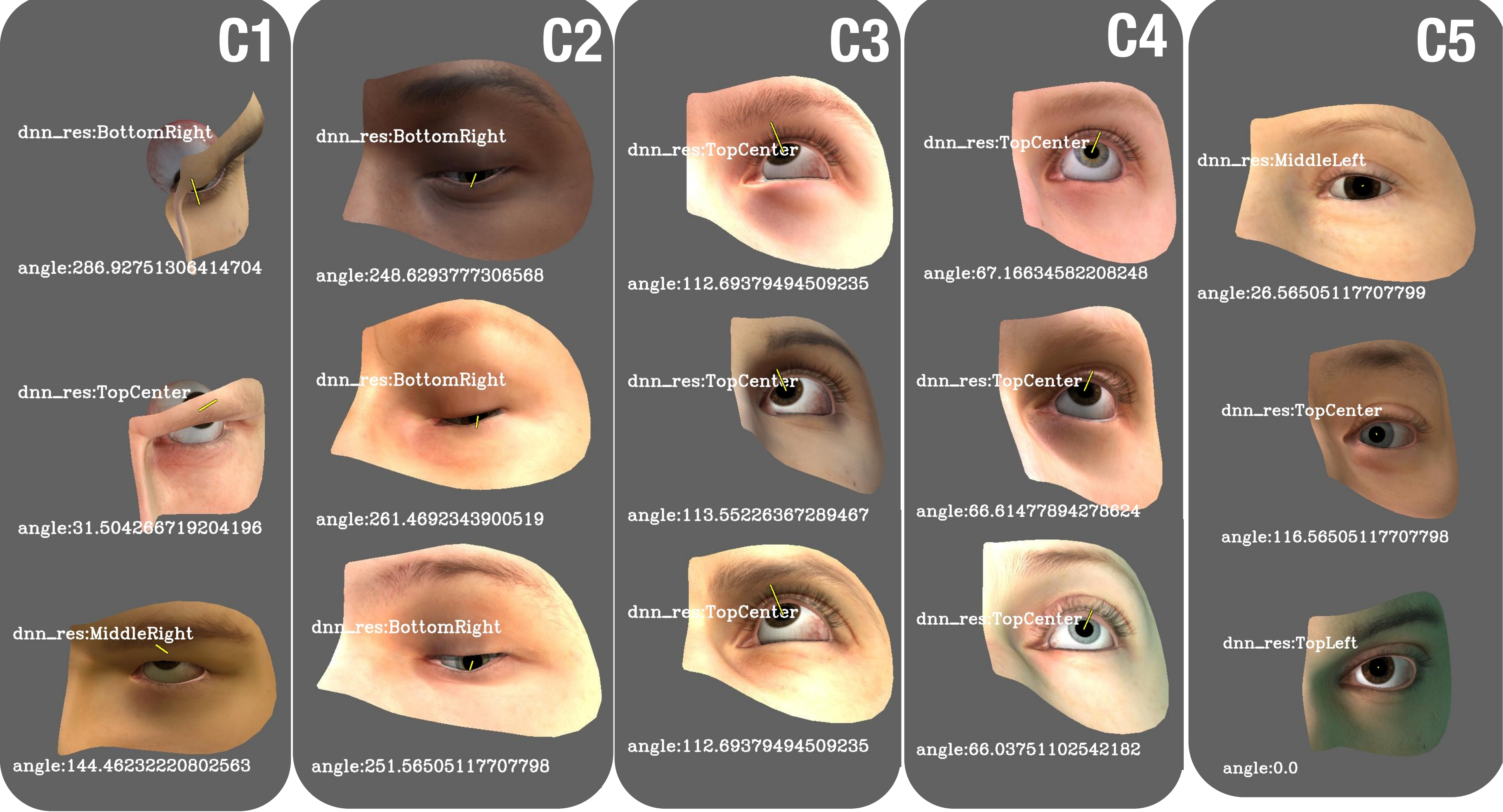}
\caption{Clustering results for \GazeDNN.}
\label{fig:ClusteringResults}
\end{figure}

Root cause clusters are then inspected by engineers to determine unsafe conditions. For example, Figure~\ref{fig:ClusteringResults} shows the clusters generated for the \GazeDNN in Figure~\ref{fig:DNNs} on a test set with eye images generated by UnityEyes. To simplify the understanding of root causes, we printed the gaze angle on each image. 
Clusters C1 and C2 group together images that lead to DNN errors because the pupil is barely visible. 
In contrast, clusters C3 and C4 group images that are misclassified because the gaze angle is close to the classification threshold.
Cluster C5, however, shows images that are misclassified because the training set labels are incomplete and do not capture the case of an eye looking middle center. 

\ASER{1.2c}{HUDD correctly handles both (1) the case in which erroneous DNN results with different output labels share the same root cause and (2) the case in which erroneous DNN results with the same output label are caused by distinct root causes.
The first case is exemplified by cluster C5, which includes images that lead to different erroneous results (e.g., TopCenter or TopLeft) due to the same root cause (i.e., the eye is looking middle center but the DNN was not trained to detect it).
The second case is exemplified by clusters C5 and C3, both including images erroneously classified as TopCenter. 
In cluster C3, this is due to the gaze angle close to the threshold with class TopRight whereas Cluster C5 includes images erroneously classified as TopCenter that are actually middle center.}

In addition, the clusters in Figure~\ref{fig:ClusteringResults} 
show that 
\ASEnn{\textbf{\APPR identifies root causes that are associated with an incomplete training set}} 
(e.g., borderline cases for gaze angle detected by C3 and C4) but also with \textbf{an incomplete definition of the predicted classes }
(i.e., the middle center gaze detected by cluster C5 and the closed eyes detected by cluster C2) and \textbf{limitations in our capacity to control the simulator} (i.e., unlikely face positions detected by cluster C1). 
\ASEnn{The first case is addressed by HUDD retraining procedures (i.e., Steps 4-7) whereas the other causes require that engineers modify the DNN (e.g., to add an output class) or improve the simulator.}

\CHANGEDNEW{To further simplify the inspection of root cause clusters, our toolset also generates a set of animated GIF images, one for each cluster~\cite{GIF}. Each generated GIF image shows all the images belonging to a cluster one after the other.
Animated GIFs enable engineers to inspect a large number of images in a few seconds (e.g., we configure our tool to visualize 100 images in a minute) thus facilitating the detection of the common characteristics among them.}

\subsection{Identification of Unsafe Images}

\APPR processes the improvement set to automatically identify potentially unsafe images.
This is done by assigning improvement set images to root cause clusters while limiting the number of assigned images.

\CHANGED{To assign images to clusters, \APPR relies on the \emph{single linkage method} (see Section~\ref{sec:back:cluster}).
According to the single linkage method, an image $y$ belonging to the improvement set $\mathit{IS}$, is assigned to the cluster containing the closest error-inducing image from the test set.
We rely on single linkage since it has a desirable property: If applied to the error-inducing test set images used to generate the clusters, it ensures that every image is assigned to the cluster it belongs to. This is important since, in a realistic scenario where test set and improvement set images are collected from the field, it ensures the selection of unsafe images that are highly similar to error-inducing ones.}

Unfortunately, single linkage alone is not sufficient to identify unsafe images. Since the root cause clusters capture only the unsafe portion of the input space, 
\CHANGEDNEW{every image in the improvement set, including the safe ones, will be assigned to a root cause cluster. In other words,} 
a safe image might be assigned to a cluster simply because it happens to be closer to an image belonging to this particular cluster.

\CHANGED{To address this problem, we heuristically select, for every cluster, the images that are likely error-inducing by estimating the number of error-inducing images for each error-cause cluster in the improvement set.
Since the improvement set is generally derived from the same population as the test set (e.g., real world images collected according to the same strategy as for the test set), we can assume that (1) the improvement set is characterized by the same accuracy as the test set and 
(2) the causes of DNN errors in the improvement set should follow the same distribution as the one observed in the test set (i.e., for every root cause cluster, we should observe a number of error-inducing images proportional to what observed in the test set).}

\CHANGED{We compute the number $U_{C_i}$ of images to be selected for a root cause cluster $C_i$, as follows:
\begin{equation}
\EquationsSize
\label{eq:U:RCC:TrS}
U_{C_i}=  (\mathit{|TestSet|} * \mathit{sf}) * (1-\mathit{TestSetAcc}) * \frac{|\mathit{C}_i|}{|\mathit{C}|}
\end{equation}}

%

%

\CHANGED{
The term $(\mathit{|TestSet|} * \mathit{sf}) * (1-\mathit{TestSetAcc})$ estimates the number of error-inducing images that should be selected from the improvement set. The term $\frac{|\mathit{C}_i|}{|\mathit{C}|}$ indicates how to distribute these images across root cause clusters in order to preserve the proportion of the test set across clusters in the improvement set.
The term $(\mathit{|TestSet|} * \mathit{sf})$ provides an upper bound for the unsafe set size as a proportion of the  test set size, which is determined by the available budget for labelling. Indeed, \CHANGED{$\mathit{|TestSet|}$ is the size of the test set, while $\mathit{sf}$ is a selection factor in the range [0-1] (we use $0.3$ in our experiments).}
The term $(1-\mathit{TestSetAcc})$ indicates the proportion of error-inducing images that should be observed in the improvement set, based on assumption (1) above. 
Multiplied by the unsafe set upper bound, it ensures that we select a fraction of it.
Finally, the term $\frac{|\mathit{C}_i|}{|\mathit{C}|}$ indicates the fraction of unsafe set images that should be assigned to the root cause cluster $\mathit{C}_i$. 
The term $|\mathit{C}_i|$ is the size of the cluster $\mathit{C}_i$. 
The term $|\mathit{C}|$ is the number of error-inducing images in the test set.
Based on assumption (2) above, the fraction $\frac{|\mathit{C}_i|}{|\mathit{C}|}$ corresponds to the proportion of error-inducing images from the test set belonging to the root cause cluster $\mathit{C}_i$. 
}

\CHANGED{Figure~\ref{algo:unsafe} shows the pseudocode of our algorithm for identifying unsafe images. 
It requires the root cause clusters $R$, the identifiers of the images belonging to the improvement set ($\mathit{IS}$),
the selection factor $sf$, the test set accuracy ($\mathit{acc}_{TS}$), the size of the test set ($\mathit{size}_{TS}$),
and a distance matrix $\mathit{DM}_{\mathit{IS}}$ with the distances between the images in $\mathit{IS}$ and the images in the error-inducing test set, for the layer selected for the identification of root cause clusters.} \TRC{To speed up the identification of unsafe images, since we focus on one specific layer, we compute the distance matrix  based on the heatmaps returned by LRP, not the normalized ones.}

\CHANGED{The algorithm works by assigning $U_{C_i}$ improvement set images to each root cause clusters $C_i$. It ensures that every image is assigned to the closest cluster $C_i$ that has not been already filled with $U_{C_i}$  images.
The algorithm also handles the presence of spurious clusters, that is, clusters that group together diverse images for which a common root cause cannot be clearly identified. Spurious clusters may be assigned with safe images or error-inducing images that should belong to other clusters. By relying on a ranking strategy for the assignment of images to clusters, we alleviate the effect of spurious clusters. When spurious clusters are fully assigned, the algorithm correctly assigns to their respective clusters all the remaining images, even if they are accidentally closer to the spurious cluster.}
%

\CHANGED{In Figure~\ref{algo:unsafe}, Lines \ref{algo:unsafe:ucStart} to \ref{algo:unsafe:ucEnd} compute, for every cluster, the value of $U_{C_i}$ according to Equation~\ref{eq:U:RCC:TrS}.
Lines~\ref{algo:unsafe:rankStart} to \ref{algo:unsafe:rankEnd}, for every image, rank clusters, based on single linkage distance. This is performed by computing the distance of the image from each cluster (Lines~\ref{algo:unsafe:rankStart:distStart}~to~\ref{algo:unsafe:rankStart:distEnd}) and then sorting clusters accordingly (Lines~\ref{algo:unsafe:rankStart:rankStart}~to~\ref{algo:unsafe:rankStart:rankEnd}), \TR{1.3}{such that the cluster in rank 1 is the closest one.}}

\CHANGED{In Lines~\ref{algo:unsafe:assignStart} to~\ref{algo:unsafe:assignEnd}, the algorithm assigns every image 
 to the closer cluster  that is not already full. It iterates over the ranks (Line~\ref{algo:unsafe:assignStart}), and for each rank $r$, it loops over the improvement set images starting from the ones that are closer to a cluster (Line~\ref{algo:unsafe:assign:closer}). If the cluster (i.e., $\mathit{clusterId}$, Line~\ref{algo:unsafe:assign:clusterId}) is not already full (Line~\ref{algo:unsafe:assign:full}), it assigns the image to the cluster (Line~\ref{algo:unsafe:assign:toCluster}). If the cluster is full (e.g., if it is a spurious cluster), the image is processed in the next iteration, i.e., when the algorithm tries to assign images to clusters ranked as  $r+1$. Note that, for each rank, every image is processed only once (Lines~\ref{algo:unsafe:assign:delete} and~\ref{algo:unsafe:assign:delete:all} delete processed images).}

\CHANGED{The algorithm returns an unsafe set with $U_{C_i}$ images selected for each cluster ${C_i}$.
The selected images are labeled by engineers when required (Step 6 in Figure~\ref{fig:Approach}) and then used for retraining.}

%




\begin{figure}

\begin{algorithmic}[1]

\scriptsize
\Require (1) $R$, root cause clusters. (2) $\mathit{IS}$, set with the identifiers of the images belonging to the improvement set. (3) $\mathit{sf}$, the selection factor used in Equation~\ref{eq:U:RCC:TrS}. (4) $size_{TS}$ the size of the test set. (5) $acc_{TS}$ the accuracy for the test set. (6) $\mathit{DM}_{\mathit{IS}}$, distance matrix capturing the distance between images in the improvement set and images in \emph{TS}. 

\Ensure an associative array with the unsafe images associated to each root cause cluster

\Statex
\Statex \textcolor{gray}{//Initialize the array that will contain all the images to be processed} 
\State $\mathit{rankedClustersPerImage}  \gets$ new associative array that will contain, 
\Statex\hspace{2cm}for every image, the IDs of clusters, 
\Statex\hspace{2cm}ranked based on their distance from the image

\Statex \textcolor{gray}{//Set the max number of images to be assigned to each cluster} 
\For{clusterID in R} \label{algo:unsafe:ucStart} 
\State $\mathit{Uc}[clusterID] \gets  (\mathit{size}_{TS} * \mathit{sf}) * (1-\mathit{acc}_{TS}) * \frac{\mathit{sizeOf}(R[\mathit{clusterID}])}{\mathit{sizeOf}(R)}$ \label{algo:unsafe:ucEnd}
\EndFor
\Statex \textcolor{gray}{//For each image, rank clusters, based on HeatmapDistance} 
\For{$\mathit{img}$ in $\mathit{IS}$} \label{algo:unsafe:rankStart} 
\Statex \textcolor{gray}{\hspace{2mm} //Generate an associative array capturing, for every cluster,} 
\Statex \textcolor{gray}{\hspace{2mm} the distance of $\mathit{img}$ from the closest image of the cluster}
\For{$\mathit{clusterID}$ in R} \label{algo:unsafe:rankStart:distStart}
\State $\mathit{clusterDists} \gets$ new associative array to store the distance of $\mathit{img}$
\Statex\hspace{2cm}from every cluster
\State $\mathit{closest} \gets$ use $DM_{IS}$ to identify the test set image that is  closer to $\mathit{img}$ 
\\\hspace{2cm}among the ones belonging to $\mathit{clusterID}$ 
\State $\mathit{clusterDists}[\mathit{clusterID}] \gets$ distance between $\mathit{closest}$ and $\mathit{img}$
\EndFor \label{algo:unsafe:rankStart:distEnd}
\Statex \textcolor{gray}{\hspace{4mm}//Put clusters in the correct rank for $\mathit{img}$} \label{algo:unsafe:rankStart:rankStart}
\For{rank in 1 .. $|R|$} 
\State $\mathit{clusterID} \gets$ position in $\mathit{clusterDists}$ containing the lowest value
\State $\mathit{rankedClustersPerImage}[rank][img] \gets$ 
\\\hspace{2cm}$<\mathit{clusterID},\mathit{clusterDists}[\mathit{clusterID}]>$
\State set $\mathit{clusterDists}[\mathit{clusterID}]$ to undefined
\EndFor \label{algo:unsafe:rankStart:rankEnd}
\EndFor \label{algo:unsafe:rankEnd}

\Statex \textcolor{gray}{//Assign images to clusters, trying to assign every image to the closer cluster first}
\For{rank in 1 .. $|R|$} \label{algo:unsafe:assignStart}
\State $img \gets$ the index $img$, in $\mathit{rankedClustersPerImage}[rank][img]$ \label{algo:unsafe:assign:closer}
\Statex\hspace{1.3cm}containing the lowest value, i.e., the image that is closer to any of 
\Statex\hspace{1.3cm}the clusters
\Statex \textcolor{gray}{\hspace{4mm}//Save the ID of the cluster that is closer to $\mathit{img}$} 
\State $\mathit{clusterId} \gets \mathit{rankedClustersPerImage}[rank][img]$ \label{algo:unsafe:assign:clusterId}
\State delete $\mathit{rankedClustersPerImage}[rank][img]$ \label{algo:unsafe:assign:delete}
\Statex \textcolor{gray}{\hspace{4mm}//Add the image to the cluster, if this is not already full}
\If{\textbf{sizeOf} ( $\mathit{unsafeSet}[\mathit{clusterId}]$ ) $< \mathit{Uc}[\mathit{clusterId}]$} \label{algo:unsafe:assign:full}
\State add $\mathit{img}$ to $\mathit{unsafeSet}[\mathit{clusterId}]$ \label{algo:unsafe:assign:toCluster}
\Statex \textcolor{gray}{\hspace{8mm}//Remove the image from the array with the images to process}
\State delete $\mathit{rankedClustersPerImage}[rank][img]$ for all the ranks \label{algo:unsafe:assign:delete:all}
\EndIf
\EndFor \label{algo:unsafe:assignEnd}

%
%
%
%

%
%

\State {\textbf{Return} $\mathit{unsafeSet}$ }

\end{algorithmic}
\caption{Algorithm for the identification of unsafe images}
\label{algo:unsafe}
\end{figure}




%
%
%

\subsection{DNN Retraining}
\label{sec:DNNretraining}

\APPR retrains the DNNs by executing the DNN training process against a data set that is the union of the original training set and the labeled unsafe set.
\APPR uses the available model to set the initial configuration for the DNN weights.
The original training set is retained to avoid reducing the accuracy of the DNN for parts of the input space that are safe (i.e., showing no error in the test set).

\APPR balances the unsafe set with bootstrap resampling~\cite{ML:Book}, i.e., it randomly duplicates the images belonging to the cluster until every cluster has the same size. 
\CHANGED{This is done to maximize the chances of eliminating every root cause of error, even the ones that are rare (i.e., the ones for which we identify less unsafe set images).}
More formally, assuming $\mathit{Max}(|U_{C_i}|)$ being the size of the largest root cause cluster, bootstrap resampling ensures that every cluster contains $\mathit{Max}(|U_{C_i}|)$ members.
The retraining process is expected to lead to an improved DNN model compared to that based on the original training set.




%
%



\section{Empirical Evaluation}
\label{sec:empirical} 

Our empirical evaluation aims to address the following research questions:

\textbf{RQ1}. Does \APPR enable engineers to identify the root causes of DNN errors?
\ASEnn{We aim to investigate}
whether images belonging to a same cluster, as generated by \APPR, present a common set of characteristics that are plausible causes of DNN errors.

\textbf{RQ2}. How does \APPR compare with traditional DNN accuracy improvement practices?
\ASEnn{We aim to investigate}
whether \APPR enables engineers to efficiently drive the retraining of a DNN compared with state-of-the-art approaches.

To perform our empirical evaluation, we have implemented \APPR as a toolset that relies on the PyTorch~\cite{PyTorch} and SciPy~\cite{SciPy} libraries.
Our toolset, case studies, and results are available for download~\cite{REPLICABILITY}.

\subsection{Subjects of the study}
\label{sec:subj}

%

\ASEnew{To address RQ1, we need to objectively and systematically identify commonalities among images belonging to the same cluster. To do so, we rely on images generated using simulators as it allows us to associate each generated image to values of the configuration parameters of the simulator. These parameters capture information about the characteristics of the elements in the image and can thus be used to objectively identify the likely root causes of DNN errors.}

We consider DNNs that implement the key features of 
gaze detection, drowsiness detection, headpose detection, and face landmarks detection systems under development at \IEE. 
The gaze detection system (hereafter referred as \GD) has been presented in Section~\ref{sec:context:case}. 
The drowsiness detection system (\CloseDNN{}) features the same architecture as the gaze detection system, except that the DNN predicts whether eyes are closed.
\CHANGED{The headpose detection system (\HPD) receives as input the cropped image of the head of a person and determines its pose according to nine classes (straight, turned bottom-left, turned left, turned top-left, turned bottom-right, turned right, turned top-right, reclined, looking up).} 
\CHANGED{The face landmark detection system (\FLD) receives as input the cropped image of the head of a person and determines the location of the pixels corresponding to 27 face landmarks delimiting seven face elements: nose ridge, left eye, right eye, left brow, right brow, nose, mouth. Each face element is delimited by several face landmarks.}

\CHANGED{\GD, \CloseDNN{}, and \HPD follow the AlexNet architecture~\cite{AlexNet} which is commonly used for image classification. \FLD, which addresses a regression problem, relies on an Hourglass-like architecture~\cite{Hourglass}. It includes 27 output neurons, each one predicting the position (i.e., pixel) of a distinct face landmark. Since a small degree of error in the detected landmarks is considered acceptable, the output of \FLD is considered erroneous if the average distance of the identified landmarks from the ground truth is above four pixels. To apply \APPR to \FLD, we generate heatmaps by backpropagating the relevance of the worst output neuron, i.e., the output neuron with the highest distance from the ground truth.
Since face elements present very different characteristics, we apply the HUDD clustering algorithm seven times, once for each face element, by selecting images whose worst output neuron corresponds to the considered face element.} 



\begin{figure}[htb]
\centering
\includegraphics[width=5cm]{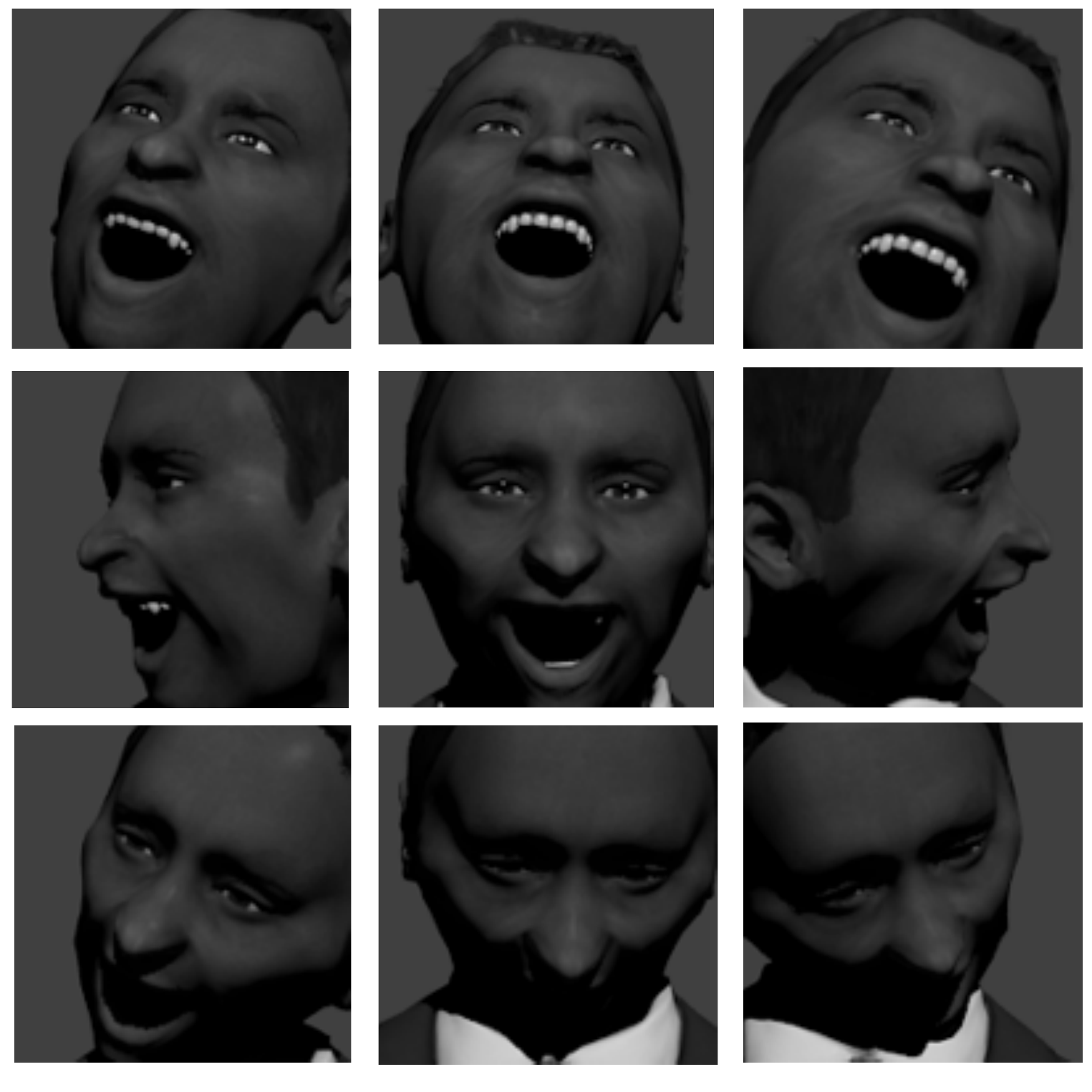}
\caption{Example of distinct head poses of the same person generated with the simulator based on MakeHuman/Blender.}
\label{fig:example:headpos}
\end{figure}

The first four rows of Table~\ref{tab:dnns} provide details about the four DNNs described above. 
Column \emph{Data Source} reports the name of the simulator generating the images used to train and test the network.
\GD and \CloseDNN have been trained and tested with images generated by UnityEyes.
Since classes need to be balanced in order to properly train the DNN,
for \CloseDNN, we selected a subset of images consisting of all the closed eyes and the same number of open eyes.
For \GazeDNN, this is not needed since UnityEyes selects the gaze angle according to a uniform distribution.
\CHANGED{\HPD and \FLD have been trained and tested with images generated using a simulator developed in-house by \IEE. The \IEE simulator relies on 3D face models built with the MakeHuman~\cite{MakeHuman} plug-in for Blender~\cite{Blender}.
To emulate car cameras, it generates grey images. \HPD and \FLD share the same training and test sets. To generate images, we used six face models for the training set, one for the test set. The use of different face models for training and testing emulates realistic scenarios in which the images processed in the field belong to persons different than the ones considered for training the DNN.
Figure~\ref{fig:example:headpos} shows examples of nine head poses generated with the same face model}.

\CHANGED{In Table~\ref{tab:dnns}, column \emph{Epochs} reports the number of epochs considered to train the network. All the DNNs have been trained for a number of epochs that was sufficient to achieve a training set accuracy above 80\%.}
Columns \emph{Training Set Size} and \emph{Test Set Size} report the size of the training and test sets. 
Columns \emph{Accuracy Training} and \emph{Accuracy Test} indicate the accuracy obtained by the DNN when executed against images in the training and test sets.
\CHANGEDNEW{Though training set accuracy is above 87\% for all the four DNNs, in the case of \HPD and \FLD, we observe a lower test set accuracy. This is due to the prediction task being more complex for \HPD and \FLD than for \GD and \OC; indeed, the DNNs for \HPD and \FLD are tested with images belonging to face models that are different than the ones used for training. This was not the case for \GD and \OC since UnityEyes does not provide the means to control face features and automatically selects them during simulation. In addition, in contrast to UnityEyes, 
the images generated with the \IEE simulator using different face models are likely to be more diverse. Indeed, the six face models integrated in UnityEyes capture only the face area surrounding the eye (i.e., eyelid and a portion of the nose) while the face models of the \IEE simulator capture the entire face. Consequently, when testing is based on new face models, it is more likely to lead to DNN errors in the case of \HPD and \FLD.
The number of face models considered for training \HPD and \FLD is limited to six since the definition of a face model is an expensive manual task.}

\begin{table}[tb]
\caption{Case Study Systems}
\footnotesize
\begin{tabular}{
|@{\hspace{1pt}}p{6mm}
|@{\hspace{1pt}}p{1.7cm}
|@{\hspace{1pt}}p{10mm}
|@{\hspace{1pt}}p{10mm}
|@{\hspace{1pt}}p{8mm}
|@{\hspace{1pt}}p{10mm}
|@{\hspace{1pt}}p{8mm}|}
\hline
\textbf{DNN}&		 \textbf{Data}	&\textbf{Training}& \textbf{Test} & \textbf{Epochs}	&\multicolumn{2}{c|}{\textbf{Accuracy}}\\
&		 \textbf{Source}& \textbf{Set Size} & \textbf{Set Size}	&  &\textbf{Training}&\textbf{Test}\\
\hline
\GD&	 UnityEyes & 61,063 &  132,630&  10& 96.84\% & 95.95\%\\
\CloseDNN&      UnityEyes &1,704	  &	4,232&  10& 87.38\% & 88.03\%\\
\HPD&      Blender &16,013  &	2,825&  10& 94.45\% & 44.07\%\\
\FLD&      Blender &16,013  &	2,825&   10& 88.97\% & 44.99\%\\
\ODDNN& CelebA~\cite{Liu:15}&7916&5276&13&83.67\%&84.11\%\\
\TrafficDNN& TrafficSigns~\cite{TRAFFICdataset}&29,416&12,631&12&92.64\%&81.65\%\\
\hline
\end{tabular}

\label{tab:dnns}
\end{table}%




Since \APPR can be applied to DNNs trained using  simulator or real images, to address RQ2, which concerns the improvement achieved after retraining the DNN, we also considered additional DNNs trained using real-world images. 
\ASEnew{We selected DNNs implementing 
traffic sign recognition (\TrafficDNN), and object detection (\ODDNN), which are typical features of automotive, DNN-based systems.} 
They are reported in the last two rows of Table~\ref{tab:dnns}. 
\TrafficDNN recognizes traffic signs in pictures.
\ASEnew{\ODDNN determines if a person wears eyeglasses. \ODDNN has been selected to compare results with MODE, a state-of-the-art retraining approach whose implementation is not available (see Section~\ref{sec:rq2}), but which is close in objective to \APPR.
\ODDNN has been trained on the same dataset used for evaluating MODE but we selected a subset of the available images to balance classes (common practice). Though the original trained model is not available, we achieved the same accuracy as the one reported. 
The other two case studies considered in the MODE evaluation were discarded because they are either not representative (i.e., low accuracy) or lack information for enabling replication (i.e., description of inputs and outputs).}
\ASEnew{\TrafficDNN and  \ODDNN follow the AlexNet architecture~\cite{AlexNet}.}


\subsection{Measurements and Results}
\label{sec:emp:clusters}

\ASEnew{We refine RQ1 into three complementary subquestions (i.e., RQ1.1, RQ1.2, and RQ1.3),
which are described in the following, along with the results obtained.}

\subsubsection{RQ1.1} \emph{Is the visual inspection of root cause clusters practically feasible?}
\label{sec:evaluation:RQ1.1}

\CHANGED{\emph{Design and measurements.} We discuss whether the number of clusters generated by HUDD is small enough to make visual inspection feasible.}

\CHANGED{Since this research question does not concern the quality of the generated clusters, we considered all the case studies, including the ones trained and tested with real-world images. For each case study system, we thus report the number of root cause clusters generated by \APPR. Also, under the assumption that engineers visually inspect five images for each root cause cluster, we discuss the ratio of error-inducing images that should be visually inspected when relying on \APPR. This ratio provides an indication of the time saved with respect to current practice (i.e., manual inspection of all the error-inducing images). A user study concerning the time savings introduced by \APPR is part of our future work.}

\emph{Results.}

\CHANGED{Table~\ref{table:RQ1.0:results} shows, for each case study, the total number of error-inducing images belonging to the test set, the number of root cause clusters generated by \APPR, and the ratio of error-inducing images that should be visually inspected when using \APPR.}

\CHANGED{For the respective DNNs, \APPR identifies 16 (\GD), 14 (\OC), 17 (\HPD), 71(\FLD), 14 (\ODDNN), and 20 (\TrafficDNN) root cause clusters. For all the case studies except \FLD, the number of root cause clusters generated is below or equal to 20. 
Assuming that engineers inspect few images (e.g., five) for each cluster in order to determine plausible root causes, manual inspection based on \APPR appear to be practically feasible. In the case of \FLD, the larger number of clusters is due to the identification of distinct root cause clusters for each face element; on average, we derive ten root cause clusters per element \TR{2.2}{(within a [6 - 11] range)}. \IEE engineers agreed that the manual inspection of a larger number of clusters is justified given the complexity of the case study.}

\CHANGED{In general, the ratio of error-inducing images that is inspected with \APPR is low, ranging from 1.49\% (\GD) to 22.84\% (\FLD), which shows that the analysis supported by \APPR saves a great deal of effort with respect to current practice (i.e., manual inspection of all the error-inducing images).}

\begin{table}[tb]
\footnotesize
\caption{Root cause clusters generated by \APPR.}

\begin{tabular}{
|@{\hspace{1pt}}p{1.2cm}
|@{\hspace{1pt}}p{20mm}
|p{20mm}
|p{20mm}|
}
\hline
\textbf{Case study}
&\textbf{\# Error-inducing images}&\textbf{\# Root cause clusters}&\textbf{Ratio of inspected images}
\\
\hline
\GD&5371&16&1.49\%\\
\OC&506&14&13.82\%\\
\HPD&1580&17&5.38\%\\
\FLD&1554&71&22.84\%\\
\ODDNN&838&14&8.35\%\\
\TrafficDNN&2317&20&4.31\%\\
\hline
\end{tabular}\\
\label{table:RQ1.0:results}
\end{table}%

\subsubsection{RQ1.2} \emph{Do the clusters generated by \APPR show a significant reduction in the variance of simulator parameters?}

\vspace{1mm}
\emph{Design and measurements.} This research question assesses if images belonging to the same cluster present similar characteristics. 
\CHANGED{To address this research question, we rely on case studies trained and tested with simulator images. 
When images are generated with a simulator, images belonging to the same cluster should present similar values for a subset of the simulator parameters.}
In turn, this should result in a reduction of variance for these parameters in comparison to the entire error-inducing test set.
For a cluster $C_{i}$, the rate of reduction in variance for a parameter $p$ can be computed as follows:

\begin{math}
RR^{p}_{C_{i}}= 1 - \frac { \mathit{variance}\ \mathit{of} p\ \mathit{for}\ \mathit{the}\ \mathit{images}\ \mathit{in}\ C_{i} } {\mathit{variance}\  \mathit{of}\  p\  \mathit{for}\ \mathit{the}\ \mathit{entire}\ \mathit{error-inducing}\ \mathit{set}}
\end{math}

\ASE{Positive values  for $RR_{C_{i}}^p$ indicate reduced variance.}


Table~\ref{tab:parameters} provides the list of parameters considered in our evaluation. 
\CHANGED{In the case of \GD and \OC,} we selected all the parameters provided by the simulator except the ones that capture coordinates of single points used to draw the pictures (e.g., eye landmarks) since these coordinates alone are not informative about the elements in the picture.
However, we considered these coordinates to compute metrics that capture information about the scene in the image. We refer to such metrics as derived parameters. 
For example, we compute the distance between the bottom of the pupil and the bottom eyelid margin (\emph{PupilToBottom} in Table~\ref{tab:parameters}). 
\ASE{It} determines if the eye is in an unusual position, e.g., if the eye is at the bottom of the orbit. 
\CHANGED{In the case of \HPD, similarly to \GD and \OC, we considered the parameters provided by the simulator, excluding once again landmark coordinates. For parameters expressed with X-Y-Z coordinates, we considered the coordinate on each axis as a separate parameter.} 
\CHANGED{In the case of \FLD, since a DNN error may depend on the specific shape and expression of the face being processed (i.e., on the specific position of a landmark), we considered the coordinates of the 27 landmarks on the horizontal and vertical axes as distinct parameters (54 parameters in total).}

We compute the percentage of clusters showing reduction in variance for at least one of the parameters. 
Since we do not know a priori the number of parameters that capture common error causes, we consider variance reduction in one parameter to be sufficient.
\ASEnew{More precisely, we compute the percentage of clusters 
with a reduction in variance between 0.0 and 0.9, with incremental steps of 0.10.}
To answer positively our research question, a high percentage of the clusters should show a reduction in variance for at least one of the  parameters.

\begin{table}[tb]
\caption{Image parameters considered to address RQ1.1}
\footnotesize
\begin{tabular}{
|@{\hspace{1pt}}p{7mm}
|@{\hspace{1pt}}p{1.5cm}
|@{\hspace{1pt}}p{5.9cm}|}
\hline
\textbf{DNN}&\textbf{Parameter}&\textbf{Description}\\
\hline
\multirow{14}{*}{GD/OC}&Gaze Angle&Gaze angle in degrees.\\
\cline{2-3}
&Openness&Distance between top and bottom eyelid in pixels.\\
\cline{2-3}
&H\_Headpose&Horizontal position of the head (degrees)\\
\cline{2-3}
&V\_Headpose&Vertical position of the head (degrees)\\
\cline{2-3}
&Iris Size&Size of the iris.\\
\cline{2-3}
&Pupil Size&Size of the pupil.\\
\cline{2-3}
&PupilToBottom&Distance between the pupil bottom and the bottom eyelid margin. 
\\
\cline{2-3}
&PupilToTop&Distance between the pupil top and the top eyelid margin. 
\\
\cline{2-3}
&DistToCenter&Distance between the pupil center of the iris center. When the eye is looking middle center, this distance is below 11.5 pixels.\\
\cline{2-3}
&Sky Exposure&Captures the degree of exposure of the panoramic photographs reflected in the eye cornea.\\
\cline{2-3}
&Sky Rotation&Captures the degree of rotation of the panoramic photographs reflected in the eye cornea.\\
\cline{2-3}
&Light&Captures the degree of intensity of the main source of illumination.\\
\cline{2-3}
&Ambient&Captures the degree of intensity of the ambient illumination.\\
\hline
\multirow{6}{*}{HPD}&
Camera Location& Location of the camera, in X-Y-Z coordinate system.\\
\cline{2-3}
&Camera Direction& Direction of the camera (X-Y-Z coordinates).\\
\cline{2-3}
&Lamp Color& RGB color of the light used to illuminate the scene.\\
\cline{2-3}
&Lamp Direction& Direction of the illuminating light (X-Y-Z coordinates).\\
\cline{2-3}
&Lamp Location& Location of the source of light (X-Y-Z coordinates).\\
\cline{2-3}
&Headpose& Position of the head of the person (X-Y-Z coordinates). It is used to derive the ground truth.\\
\hline
\multirow{2}{*}{FLD}&
X coordinate of landmark& Value of the horizontal axis coordinate for the pixel corresponding to the $i^{th}$ landmark.\\
\cline{2-3}
&Y coordinate of landmark& Value of the vertical axis coordinate for the pixel corresponding to the $i^{th}$ landmark.\\
\hline
\end{tabular}

\label{tab:parameters}
\end{table}%






\begin{figure}[htb]
\includegraphics[width=8cm]{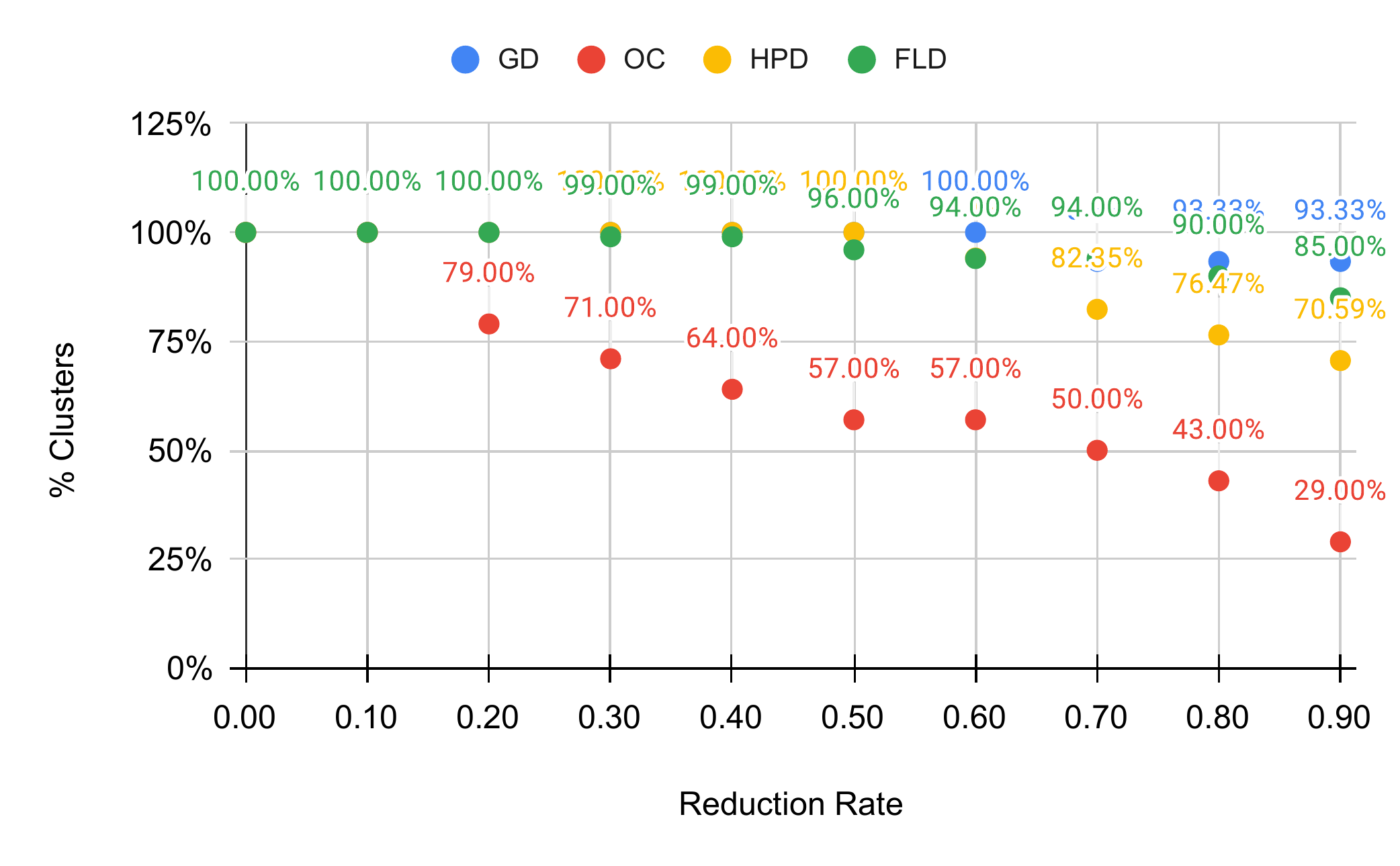}
\caption{RQ1.1: Clusters with at least one parameter showing a reduction rate above thresholds in the range (0.0 - 1.0).}
\label{fig:RQ1.1:cl}
\end{figure}

\emph{Results.} Figure~\ref{fig:RQ1.1:cl} shows the percentage of clusters with variance reduction for at least one of the simulator parameters, at different reduction rates.

\CHANGED{We can positively answer RQ1.1 since all the clusters present at least one parameter with a positive reduction rate ($>0$ in Figure~\ref{fig:RQ1.1:cl}).
Also, a very high percentage of the clusters (i.e., 57\% for \CloseDNN, 96\% for \FLD, and 100\% for both \GD and \HPD) include at least one parameter with a reduction rate above or equal to 0.5, i.e., 50\% reduction in variance.}
Expectedly, as the threshold considered for variance reduction increases, the percentage of clusters tends to decrease. 
\CHANGED{However, a 0.9 threshold is still matched by 29\% (\OC) to 93.33\% (\GD) of the clusters (69.48\% on average), a very high proportion, thus showing that most of the clusters should present a noticeable common characteristic.}

\begin{table}[tb]
\caption{Safety parameters considered to address RQ1.3}
\footnotesize
\begin{tabular}{
|@{\hspace{1pt}}p{7mm}
|@{\hspace{1pt}}p{1.6cm}
|@{\hspace{1pt}}p{5.8cm}|}
\hline
\textbf{DNN}&\textbf{Parameter}&\textbf{Unsafe values}\\
\hline
\multirow{7}{*}{\GD,\OC}&Gaze Angle&Values used to label the gaze angle in eight classes (i.e., 22.5$^{\circ}$, 67.5$^{\circ}$, 112.5$^{\circ}$, 157.5$^{\circ}$, 202.5$^{\circ}$, 247.5$^{\circ}$, 292.5$^{\circ}$, 337.5$^{\circ}$).\\
&Openness&Value used to label the gaze openness in two classes (i.e., 20 pixels) or an eye abnormally open (i.e., 64 pixels).\\
&H\_Headpose&Values indicating a head turned completely left or right (i.e., 160$^{\circ}$, 220$^{\circ}$)\\
&V\_Headpose&Values indicating a head looking at the very top/bottom (i.e., 20$^{\circ}$, 340$^{\circ}$)\\
&DistToCenter&Value below which the eye is looking middle center (i.e., 11.5 pixels).\\
&PupilToBottom&Value below which the pupil is mostly under the eyelid (i.e., -16 pixels).\\
&PupilToTop&Value below which the pupil is mostly above the eyelid (i.e., -16 pixels).\\
\hline
\multirow{2}{*}{\HPD}&Headpose-X&Boundary cases (i.e.,-28.88$^{\circ}$,21.35$^{\circ}$), values used to label the headpose in nine classes (-10$^{\circ}$,10$^{\circ}$), and middle position (i.e., 0$^{\circ}$).\\
\cline{2-3}
&Headpose-Y&Boundary cases (i.e.,-88.10$^{\circ}$,74.17$^{\circ}$), values used to label the headpose in nine classes (-10$^{\circ}$,10$^{\circ}$), and middle position (i.e., 0$^{\circ}$).\\
\hline
\end{tabular}

\label{tab:boundary}
\end{table}%




\begin{table*}[t!]
\footnotesize
\caption{RQ2. Size of Images Set used for Retaining and Accuracy Improvement}

\begin{tabular}{
|@{\hspace{1pt}}p{7mm}@{\hspace{1pt}}
|@{\hspace{1pt}}p{4mm}p{3mm}p{6mm}
|@{\hspace{1pt}}p{5mm}@{\hspace{5pt}}p{5mm}@{\hspace{7pt}}p{6mm}
|@{\hspace{1pt}}p{5mm}@{\hspace{7pt}}p{5mm}
|p{4mm}
|p{20mm}@{\hspace{1pt}}
|@{\hspace{1pt}}p{20mm}@{\hspace{1pt}}
|@{\hspace{1pt}}p{20mm}@{\hspace{1pt}}
|p{8mm}@{\hspace{1pt}}
|p{5mm}@{\hspace{3pt}}
p{6mm}@{\hspace{1pt}}
|
}
\hline
&\multicolumn{8}{c}{\textbf{Size of Images Sets for Retraining}}
&\multicolumn{1}{|c|}{\textbf{Accuracy}}
&\multicolumn{3}{c|}{\textbf{Accuracy (Accuracy improvement)}}
&\multicolumn{1}{c|}{\textbf{Delta}}
&\multicolumn{2}{c|}{\textbf{$\hat{A}_{12}$}}

\\
&\multicolumn{3}{c}{\textbf{HUDD}}
&\multicolumn{3}{c}{\textbf{B1}}&\multicolumn{2}{c|}{\textbf{B2}}
&\multicolumn{1}{|c|}{\textbf{Original}}
&\multicolumn{3}{c|}{}
&\multicolumn{1}{c|}{\textbf{wrt best}}
&\multicolumn{2}{c|}{\textbf{HUDD vs}}
\\
\textbf{DNN}
&\textbf{IS}&\textbf{US}
&\textbf{BLUS}
&\textbf{IS}&\textbf{LUS}&\textbf{ALUS}
&\textbf{IS}&\textbf{ALIS}
&\multicolumn{1}{|c|}{\textbf{Model}}
&\textbf{HUDD}&\textbf{B1}&\textbf{B2}
&\multicolumn{1}{c|}{\textbf{Baseline}}
&\textbf{B1}&\textbf{B2}
\\
\hline
\GD
&72500
&1615
&4192	
&1615
&156.4
&4192
&1615
&4192
&95.95\%
&96.23\% (+0.28)
&95.77\% (-0.18)
&95.80\% (-0.15)
&+0.43\%
&0.72
&0.71
\\
\CloseDNN&
4103&
160&
336&
160&
43.4&
336&
160&
336&
88.03\%&
94.41\% (+6.38)&
91.65\% (+3.62)&
92.33\% (+4.30)&
+2.08\%&
1.00&
1.00
\\
\HPD&
4700&
481&
697&
481&
12.7&
697&
481&
697&
44.07\%&
68.13\% (+24.06)&
66.73\% (+22.66)&
66.30\% (+22.23)&
+1.40\%&
0.70&
0.72
\\
\FLD&
6864&
502&
970&
502&
34.9&
970&
502&
970&
44.99\%&
75.23\% (+30.24)&
72.02\% (+27.03)&
73.83\% (+28.84)&
+1.40\%&
0.70&
0.60
\\
\TrafficDNN
&9775
&704
&1860
&704
&53.2
&1860
&704
&1860
&81.65\%
&93.03\% (+11.38)
&92.63\% (+10.98)
&92.73\% (+11.08)
&+0.30\%
&0.83
&0.67
\\
\ODDNN
&13194
&258
&770
&258
&69.1
&770
&258
&770
&84.12\%
&97.04\% (+12.92)
&96.63\% (+12.51)
&96.67\% (+12.55)
&+0.37\%
&0.79
&0.60
\\
\hline
\end{tabular}\\
IS: Improvement Set, US: Unsafe Set, 
BLUS: Balanced Labeled Unsafe Set;
LUS: Labeled Unsafe Set (average over all the runs);
ALUS/ALIS: Augmented Labeled Unsafe/Improvement Set.
\label{table:RQ2.2:details}
\end{table*}%

\subsubsection{RQ1.3} \emph{Do parameters with high reduction in variance identify the plausible cause for DNN errors? }

\vspace{1mm}
\emph{Design and measurements.}
With RQ1.3, we ask whether the commonalities of the images belonging to the root cause clusters can help engineers determine the root causes of the DNN errors.

We expect DNN errors to be triggered in specific parts of the input space, each one capturing characteristics of the input images.
To identify the input sub-spaces that are unsafe for our case studies,
based on domain knowledge, we have identified a set of parameters (hereafter, \emph{unsafe parameters}) 
for which it is possible to identify values (hereafter, \emph{unsafe values}) around which, or below which, we are likely to observe a DNN error. 
\CHANGED{However, for \FLD, it was not possible to determine, a priori, a set of unsafe parameters that might affect the results and we had to leave out that case study.}
\CHANGEDNEW{Indeed, the position of a landmark may depend on many factors including the shape of the face element (e.g., thick lips), the element position (e.g., mouth being open), the headpose, and the camera position. Since the \IEE simulator does not export information about the shape and position of face elements, it was not possible to define a set of metrics capturing plausible error causes}. 

Table~\ref{tab:boundary} provides the list of unsafe parameters, along with the unsafe values identified.
For example, for the Gaze Angle parameter, unsafe values consist of the boundary values used to label images with the gaze direction.

Root cause clusters that are explanatory should present at least one characteristic that is noticeable by the engineer, i.e., they should have at least one parameter with high (i.e., 50\%) reduction in variance.
In addition, 
at least one of the parameters with high variance reduction should be an unsafe parameter.
Finally, the cluster average should be close to one unsafe value. 
For Gaze Angle, Openness, H\_Headpose, V\_Headpose, \CHANGED{Headpose-X, and Headpose-Y,}
since unsafe values split the parameter domains into subranges,
we determine that the cluster average is close to one unsafe value  
if the difference between them 
is below 25\% of the subrange including the average value.
For DistToCenter, PupilToBottom, and PupilToTop, we simply check if the average is below or equal to the unsafe value.
Finally, we compute the percentage of clusters for which the conditions above hold.
To answer positively to RQ1.3, this percentage should be high.

\vspace{1mm}
\emph{Results.} 
\CHANGED{In the case of \GD, according to the conditions defined above, the percentage of clusters that identify the likely root cause of DNN errors is very high: 86.66\% 
(13 out of 15). The identified unsafe parameters are \emph{Angle}, \emph{Openness}, and \emph{DistToCenter}. 
For one cluster not meeting the conditions, the unsafe parameters (i.e., \emph{DistToCenter}) have a reduction in variance of 44\%, below the 50\% threshold. This threshold is, however, arbitrary and a manual inspection of the cluster clearly shows that the commonality is the eye being abnormally open.
The other non-compliant cluster shows pupils being partially masked by the eyelid; however, we could not define a measure to systematically capture this situation based on simulator parameters.}

\CHANGED{For \CloseDNN, we obtain 57.14\% (8 out of 14), with \emph{Openness} and \emph{X\_Headpose} being the unsafe parameters.
The remaining clusters are characterized either by thin almond eyes, an aspect of the simulation that is not controllable with parameters, and pupils being partially masked by the eyelid.}

\CHANGED{In the case of \HPD, we obtain 88.24\% (15 out of 17). For the two remaining clusters, the common characteristic is the presence of visible white teeth, which are not visible in training set images and may confuse the DNN since they stand out in grey-scale images. 
Based on the above observations, we respond positively to RQ1.3 since, in all cases, clusters are clearly associated with image characteristics that are plausible causes of errors.}

\subsubsection{RQ2} \emph{How does \APPR compare to traditional DNN accuracy improvement practices?}
\label{sec:rq2}

\begin{figure}[b]
\vspace{-4mm}
\includegraphics[width=8cm]{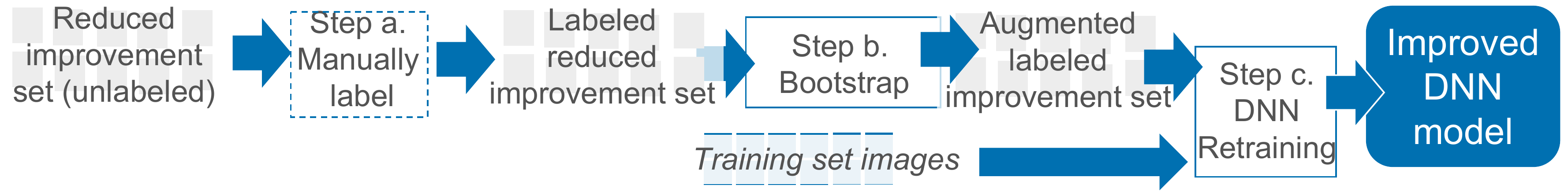}
\caption{Baseline 2 (B2).}
\label{fig:Baseline2}
\end{figure}


This research question aims to compare the accuracy improvements achieved by \APPR with the improvements achieved by baseline approaches, which do not rely on the automated selection of predicted unsafe images.

We consider two baseline approaches, namely B1 and B2.
B1 has been introduced in Section~\ref{sec:context:debugging} and consists of selecting for retraining the misclassified images belonging to the labeled improvement set.  
\ASEnn{B2 is depicted in Figure~\ref{fig:Baseline2}. It follows the HUDD process except that it selects unsafe images randomly (i.e., the \emph{Reduced improvement set}) instead of relying on root cause clusters. B2 enables the evaluation of the benefits of selection based on root cause clusters over random selection.} 

To not introduce bias in the results, we rely on the same experiment setting for all the approaches (i.e.,
same configuration of the DNN training algorithm and 
same number of images to be labeled). 
In the case of \APPR, only the images in the unsafe set need to labeled.
In the other cases, all the images in the improvement set must be labeled. 
For this reason, for the two baselines, we select an improvement set
that is a random subset of the improvement set used by \APPR (referred to as \emph{reduced improvement set}) 
and has the same size as the unsafe set generated by \APPR. 
To account for randomness, we repeat the experiment 10 times.

With \APPR, retraining the DNN was done by applying the approach described in Section~\ref{sec:DNNretraining}.
\ASEnn{For B1 and B2, we configure bootstrap resampling to generate an \emph{augmented labeled unsafe set} and an \emph{augmented labeled improvement set} with the same size as the \emph{balanced labeled unsafe set} for \APPR.}

To answer the research question, we compute the accuracy of the retrained models on the test set and compare the accuracy improvement obtained by \APPR with that obtained by the baselines.
We considered all the case studies listed in Table~\ref{tab:dnns}. 
The improvement set for \GD and \CloseDNN has been generated  through additional executions of UnityEyes.
\CHANGED{To simulate a realistic scenario in which engineers collect additional data from the field or construct additional simulator models to improve DNN accuracy, the improvement sets for \HPD and \FLD have been generated with additional executions of the IEE simulator configured to use two new face models, which were not used for generating the training and test sets.}
\ASEnn{For the other cases, we selected images of the original datasets which had not been used for the training and test sets.}
\emph{Results.} The first eight columns of
Table~\ref{table:RQ2.2:details} provide the number of images used to retrain the DNNs. 
%
%
The remaining columns of Table~\ref{table:RQ2.2:details} show the accuracy of the retrained models, the delta with respect to the best baseline, and $\hat{A}_{12}$ effect size~\cite{VDA}. 
For the accuracy, negative values indicate that the accuracy of the retrained model is worse than that of the original model.
\APPR always fares better than the baseline approaches. 
\ASEnn{Vargha and Delaney’s $\hat{A}_{12}$ effect size is always equal or above 0.60, which indicates that, in all cases, \APPR has 
higher chances of generating accurate DNNs than baselines~\cite{VDA,Arcuri:StatisticalGuide:2011}. In the paper by Vargha and Delaney~\cite{VDA}, the authors specify the an effect size above 0.56 suggests a significant difference, with higher thresholds for medium (0.64) and large (0.71) effects.}

\CHANGED{\APPR accuracy improvements range from 0.28\% to 30.24\%.} In contrast, B1 and B2 improvements range from -0.18\% to 27.03\% and -0.15\% to 28.84\%, respectively.
For DNNs with an accuracy above 80\%, we have the following ranges: from 0.28\% to 12.92\% (HUDD), from -0.18\% to 12.51\% (B1), and -0.15\% to 12.55\% (B2).
For \HPD and \FLD, which have lower initial accuracy, we have the following ranges: from 12.92\% to 24.06\%  (HUDD), from -0.18\% to 12.51\% (B1), and -0.15\% to 12.55\% (B2). We can therefore conclude that \APPR is most useful when DNNs have lower accuracy and there is more room for improvement. Those are also the cases where retraining is most particularly important. 
The negative results obtained by the baselines for \GD suggest that retraining the DNN without targeting the DNN-error root causes may lead to worse accuracy.
The choice of an inadequate strategy for retraining DNNs is therefore particularly detrimental since one could invest significant time and effort in labeling improvement set images without getting any benefit.


The difference in accuracy improvement between \APPR and the best baseline ranges between 0.30 (\TrafficDNN) and 2.08 (\CloseDNN).
Given that all techniques cost the same according to our experiment design, it is therefore recommended to use \APPR. 
\CHANGED{In addition, there is a larger average difference ($\ge 1.40$) between \APPR and baselines in the cases of \OC, \HPD, and \FLD. There are three plausible reasons to explain these differences. One is that, for \HPD and \FLD, there is significant room for accuracy improvement in the original models. Second, to increase the accuracy of \HPD and \FLD, we require improvement set images that are very different from the training set ones (i.e., pictures generated with new face models). We deem this to be a realistic situation that generalizes beyond our case studies; for example, it might be observed also in autonomous driving systems where certain types of vehicles (e.g., e-scooters) are missing from the training set.}
\CHANGEDNEW{Third, for the three DNNs above, the training set is missing unsafe situations where the predictions are expected to be challenging (e.g., very dim light in the image). 
The type of retraining described above is particularly important in our context since the reduction of the unsafe input space is a key objective of safety engineering practices for the automotive industry~\cite{SOTIF}.}

\ASEnn{Though the above accuracy differences may appear small, they may nevertheless be important in the context of critical applications where every percentage point in improvement matters. Furthermore, one should recall that, when we are dealing with highly accurate DNNs, room for improvement is limited. 
}

\ASEnew{Finally, results with \ODDNN show that \APPR achieves better accuracy than MODE (97.04\% vs 89\%~\cite{Ma2018} after DNN retraining). These results show the potential of \APPR which, in addition to a higher accuracy than MODE, also provides root cause clusters.}

\subsection{Execution Time} 

\begin{table}[tb]
\caption{Experiments Execution Time}
\footnotesize
\begin{tabular}{
|@{\hspace{1pt}}>{\raggedleft\arraybackslash}p{6mm}@{\hspace{1pt}}|
@{\hspace{1pt}}>{\raggedleft\arraybackslash}p{10mm}@{\hspace{1pt}}|
@{\hspace{1pt}}>{\raggedleft\arraybackslash}p{9mm}@{\hspace{1pt}}|
@{\hspace{1pt}}>{\raggedleft\arraybackslash}p{9mm}@{\hspace{1pt}}|
@{\hspace{1pt}}>{\raggedleft\arraybackslash}p{8.5mm}@{\hspace{1pt}}|
@{\hspace{1pt}}>{\raggedleft\arraybackslash}p{8.5mm}@{\hspace{1pt}}|
@{\hspace{1pt}}>{\raggedleft\arraybackslash}p{8.5mm}@{\hspace{1pt}}|
@{\hspace{1pt}}>{\raggedleft\arraybackslash}p{12mm}@{\hspace{1pt}}|
>{\raggedleft\arraybackslash}p{6mm}@{\hspace{1pt}}|
}
\hline
\textbf{DNN}&	\textbf{Training}&	\textbf{Testing}&	\textbf{\APPR}& \textbf{\APPR} & 	\textbf{BL1}& \textbf{BL2}& \textbf{Testing} &\textbf{Total}\\
&	&	&	\textbf{Step} & \textbf{Steps}  & &	& \textbf{Improved} & \\
&	&	&	\textbf{1}& \textbf{4-7}& &	& \textbf{DNNs} & \\
\hline
\GD	& 
2.66\% &	2.00\% &	3.82\% &	26.91\% &	31.97\% &	26.64\% &	5.99\% &
375.3 \\

\CloseDNN	&
1.34\% &	8.93\% &	18.30\% &	15.63\% &	15.63\% &	13.39\% &	26.79\% &
3.7 \\

\HPD	& 
2.01\% &	2.41\% &	21.16\% &	22.93\% &	24.14\% &	20.11\% &	7.24\% &
20.7\\
\FLD &	
2.81\% &	1.41\% &	2.81\% &	28.43\% &	35.60\% &	28.10\% &	0.84\% &
177.9\\

\ODDNN	& 
2.80\% &	0.73\% &	6.83\% &	29.13\% &	30.81\% &	28.01\% &	1.68\% &
29.8\\
\TrafficDNN&	
1.89\% &	1.94\% &	29.36\% &	20.47\% &	22.63\% &	18.86\% &	4.85\% &
30.9\\
\hline
\end{tabular}

\label{tab:time}
\end{table}%



\TR{2.1}{Table~\ref{tab:time} provides details about the time required to perform our experiments. It reports on the total execution time and how it is distributed across the different steps of \APPR and the execution of the baseline approaches. 
In Table~\ref{tab:time}, columns 5 to 7 refer to the cumulative time over 10 repetitions, column 8 refers to the cumulative time required for testing the retrained models for \APPR, BL1, and BL2. Our experiments took between 3.7 (OC) and 375.3 hours (GD) across DNNs, which highlights the large endeavor entailed by repeating experiments ten times in order to be able to draw statistical conclusions. 
Execution time is driven by the size of the data sets (i.e., executions with DNNs trained and tested with larger data sets took more time). The time required by \APPR to improve the DNN (i.e., \emph{Steps 4-7})--- which includes also the identification of unsafe images---is similar with the time required by baseline approaches (columns 6 and 7), thus showing that \APPR does not introduce delays in the retraining process. The time required to perform \APPR Step 1 is significant (between 40 minutes and 14 hours); however, Step 1 can be executed overnight and help reduce human effort to identify root cause clusters.}

\TRC{To be able to execute experiments for 638.3 hours, we parallelized the executions of experiments using the HPC cluster of the University of Luxembourg~\cite{HPCS14}. We relied on Intel Xeon Gold 6132 nodes (2.6 GHz with 
four Tesla V100 16G SXM2).}

\subsection{Threats to validity} 

We target DNNs performing image analysis in the perception layer of safety-critical systems. 
To address threats to external validity, for RQ2, we have considered DNNs performing classification of body parts and road objects, which are typical features in automotive systems. Further, one regression DNN was also analyzed, thus showing the applicability of the approach beyond classification; to this end, we considered a DNN representative of the typical task of landmark detection. 
\TR{2.3}{Though four subject DNNs out of six  implement tasks motivated by IEE business needs, they address problems that are quite common in the automotive industry (i.e., angle determination and landmarks detection). To strengthen  generalizability,  for two of these four case studies (i.e., GD and OC), we relied on a third-party simulator used in computer vision research (i.e., UnityEyes). Further, we considered two case studies from related work (i.e., TSR and OD) that rely on real images. Our benchmark DNNs are therefore both diverse and representative.}

For RQ1, we could only consider a subset of the case studies having high-resolution simulators available. Simulation is based on Blender~\cite{Blender}, a readily available and widely used technology, thus making our results more representative. 
In our experiments, to objectively and systematically evaluate the quality of the generated clusters, we relied on the analysis of simulator parameters rather than a user study, which, however, should be undertaken in the future.
\TR{2.2}{Though the fidelity of simulator images is always a question, our experimental results do not show different trends for real and simulated images with respect to the number of clusters and accuracy improvements.
Indeed, the number of root cause clusters identified for the two DNNs working with real-world images (\ODDNN and \TrafficDNN), which are 14 and 20, are similar to the ones observed in DNNs with simulator images (ranging from 11 to 17). Also, the accuracy improvement obtained for \ODDNN and \TrafficDNN, i.e., +11.38 and +11.92 (see Table~\ref{table:RQ2.2:details}), is within the range observed with simulator images (i.e., +0.28 and +30.24).}

\TR{2.1}{In our work, we do not rely on the popular K-means algorithm because of its higher computational cost in our context (see Section~\ref{sec:back:cluster}). 
However, more computation might be justified by better performance results, which may include a higher variance reduction in root cause clusters, a larger number of explanatory root cause clusters, and higher accuracy  improvement.
Since our experiments have shown that (1) \APPR generates cohese and explanatory root cause clusters (RQ1),  (2) the room for accuracy improvement is small (RQ2 results show that, for four out of six DNNs, \APPR accuracy is above 93\%), and (3) our experiments already took 638.3 hours to complete, the empirical comparison of HAC with K-means and other clustering algorithms has not been considered a priority in this work. 
}

Though \APPR background technology (i.e., LRP and HAC) is context-independent, future work will investigate the evaluation of the approach in different contexts (e.g., space industry).

\section{Related Work}
\label{sec:related}

\ASEnnn{Most of the DNN testing and analysis approaches are summarized in recent surveys~\cite{huang2018survey,zhang2019machine}.}
However, research on the automated debugging and repair of DNNs is still at very early stages.

\ASER{1.4}{Under-approximation boxes~\cite{Pasareanu:19} consist of the minimal set of neurons, belonging to a specific layer, that ensure a postcondition (e.g., the generation of a specific DNN output). When applied to explain misclassifications, they lead to heatmap-like images showing the minimal set of input pixels leading to the same DNN result.}
\ASER{1.5}{Similarly, Ribeiro et al., identify the image chunks that are sufficient to generate a certain DNN result~\cite{Ribeiro:18}.}
\ASE{Like heatmap generation techniques, these two approaches cannot automatically identify the root cause for a group of error-inducing images but require  manual inspection for every error-inducing image.}

Decision trees can identify patterns of neuron activations common to a same output class~\cite{Pasareanu:19}. They are not used to explain misclassifications~\cite{Pasareanu:19} since they cannot 
be applied to look for 
patterns in root cause clusters which are not known a priori.

MODE automatically identifies the images to be used to retrain a DNN~\cite{Ma2018}. 
However, it cannot identify the root causes of DNN errors, which is a major limitation in our context.
\APPR and MODE differ also regarding the selection of images to be used for retraining, which, in the case of MODE, 
is not based on heatmaps but on training additional DNN layers that capture commonalities among neuron activations leading to DNN errors. MODE, therefore, entails repeated modification and retraining of the DNN under test, just to select the improvement set, which is a very expensive endeavor. 

\ASE{Surprise adequacy measures the degree of variation in neuron activations between a new image and the training images belonging to the same class~\cite{Feldt19}. 
Empirical results show that a retraining set with a varying degree of surprise adequacy improves DNN robustness against adversarial examples. 
However, it has never been adopted to improve accuracy for non-adversarial inputs.
Also, like previous techniques, it cannot be used to identify root causes of DNN errors.}

\ASER{1.1}{DeepFault identifies a set of suspicious neurons to synthesize new, adversarial images and improve DNN adversarial robustness~\cite{DeepFault}. Since it relies on synthesized adversarial inputs, it cannot improve accuracy for unsafe, non-adversarial inputs. Once again, it does not distinguish different root causes.}

Apricot~\cite{Zhang19} repairs DNNs by changing the weights of the DNN model. 
It works by training multiple DNNs on subsets of the training and test sets. 
The repair process aims to minimize (maximize) the distance between the weights of the DNN to repair and the weights of DNNs leading to better (worse) accuracy.
\CHANGEDNEW{Unfortunately, the accuracy improvement achieved by Apricot is lower than 2\%.}

\ASEnn{Gao et al.~\cite{Gao:Sensei:ICSE:20} and Engstrom et al~\cite{Engstrom:2019} rely on image transformations (e.g., rotations) to augment the training set and improve DNN robustness, thus addressing a different problem.}

Active learning had been proposed to minimize the number of inputs that require to be labeled to train a machine learning model~\cite{Settles2012}.
State-of-the-art approaches identify the inputs for which an incrementally trained model generates uncertain results; 
for binary classification, uncertainty can be measured by means of entropy~\cite{Settles2012}.
However, traditional active learning approaches 
that identify individual images for incremental retraining are
 unsuitable for CNNs that require a large training sets.
For CNNs, a state-of-the-art approach consists of the generation of a coreset (a small summary of large data sets) with k elements~\cite{Sener2018}. It outperforms approaches based on uncertainty sampling~\cite{shen-etal-2017-deep}.
Though promising, such approach addresses a problem different than ours, i.e., minimizing the cost of training, under the assumption that labelling cost is uniform for all the images. 
In our context, thanks to the use of simulators, labelling costs concern mostly the real-world images used to test or improve the accuracy of the DNN; for this reason, \APPR focuses on the selection of images to be used for re-training.
Instead, coresets are not meant to be used to reduce an improvement set in the presence of a model that has already been trained on a large set of inputs.
Another limitation of state-of-the-art active learning solutions is that they are inapplicable with regression DNNs. Indeed, solutions for regression models incur computational costs that are cubic with respect to the number of model parameters, thus being infeasible for the DNNs used in image processing tasks~\cite{Settles2011}. 

\TR{2.4}{Other approaches aim to reduce the costs associated with the labeling of the test set by minimizing its size~\cite{Li:Boosting:2019,PACE}.
Though they address a different problem, 
Chen et al.~\cite{PACE} rely 
on clustering applied to DNN features (e.g., neuron activations).
More precisely, they use  HDBSCAN~\cite{McInnes2017} to identify an optimal configuration for the state-of-the-art DBSCAN algorithm~\cite{DBSCAN}.
Also, they demonstrate that the FastICA dimensionality reduction algorithm~\cite{FastICA} is enabling HDBSCAN to produce the best clustering results.
Though the authors point to the execution time of HDBSCAN as one of its main limitations, 
the combination FastICA and HDBSCAN should be evaluated as part of future work to improve the root cause clusters generated by HUDD.}

\ASEnn{With respect to a recent taxonomy of DNN faults~\cite{humbatova:2019}, HUDD can identify different types of \emph{training} and \emph{input} faults, while automatically addressing problems due to \emph{training data quality} (see Section~\ref{sec:rootCausesInspection}). A more extensive evaluation of HUDD based on this taxonomy is part of our future work.}

To summarize, \APPR is the first approach that facilitates the scalable identification of distinct failure root causes in DNNs, by applying clustering algorithms to heatmaps generated by DNN explanation techniques. For the latter, we rely on 
Layer-Wise Relevance Propagation (LRP),
which is 
based on theoretical foundations that are generalizable to other DNN architectures. Further, \APPR relies on standard retraining procedures based on back propagation and gradient analysis, that have been widely applied and validated, and does not entail the direct modification of the learned DNN model. 
\CHANGEDNEW{As demonstrated in our empirical evaluation, HUDD can successfully support the debugging of DNNs that implement either classification or regression tasks.}

\section{Conclusion}
\label{sec:conclusion}

%

In this paper we introduced \APPR, an approach that automatically identifies 
the different situations in which an image processing DNN is likely to produce erroneous results.
\APPR generates clusters (i.e., root cause clusters) containing misclassified input images 
sharing a common set of characteristics that are plausible causes for errors. 
This is achieved through an hierarchical agglomerative clustering algorithm applied 
to heatmaps capturing the relevance of neurons across different DNN layers on the result.

In addition, \APPR minimizes the effort required to select and label 
additional images to be used to augment the training set and improve the DNN.
This is done by automatically selecting images that are close \CHANGEDNEW{to members of root cause clusters and thus unsafe.}
Only these selected images then need to be labeled by engineers.
\ASER{3.3}{Since DNN errors are often due to an incomplete training set (e.g., lack of images with a gaze angle close to borderline), HUDD alleviates the problem by augmenting the training set with unsafe images.} 

Empirical evaluation with simulator images show that \APPR generates clusters 
of images sharing similar values for some of the simulation parameters driving the generation of images. We can conclude that such clusters can then serve as a useful instrument for the identification of root causes of DNN errors, as exemplified in our case studies. In turn, this information is important to safety analysis as it helps clearly characterize unsafe inputs, a requirement in safety standards. Our results, on both simulated and real images, also show how these clusters can be effectively used to select new images for retraining in a way that is more efficient than existing practices and leading to better DNN accuracy. 


%



\section*{Acknowledgment}
This project has received funding from IEE Luxembourg,
Luxembourg’s National Research Fund (FNR) under grant 
BRIDGES2020/IS/14711346/FUNTASY,
 the European Research Council (ERC) under the European
Union’s Horizon 2020 research and innovation programme (grant agreement No 694277), and
NSERC of Canada under the Discovery and CRC programs. 
Authors would like to thank Thomas Stifter and Jun Wang from IEE for their valuable support.

The experiments presented in this paper were carried out
using the HPC facilities of the University of Luxembourg
(see http://hpc.uni.lu).
\ifCLASSOPTIONcaptionsoff
  \newpage
\fi

\end{document}